\documentclass[prd,preprint,tightenlines, preprintnumbers,superscriptaddress,amsmath,amssymb,nofootinbib]{revtex4-1}
 \usepackage[dvips,final]{graphicx}
  \usepackage{amssymb}
   \usepackage{amsmath}
    \usepackage{amsfonts}
     \usepackage{epsfig}
      \usepackage{bm}% bold math
\usepackage{appendix}
\usepackage[section]{placeins}

\usepackage{sidecap}
\usepackage{multirow}
\usepackage{booktabs}
\usepackage{array}
\usepackage{tabularx}
\usepackage{xcolor}
\usepackage{pstricks}

\begin{document}

\newcommand{\ds}{\displaystyle}
\newcommand{\mc}{\multicolumn} 
\newcommand{\bce}{\begin{center}}
\newcommand{\ece}{\end{center}}
\newcommand{\beq}{\begin{equation}}
\newcommand{\eeq}{\end{equation}}
\newcommand{\bea}{\begin{eqnarray}}

\newcommand{\eea}{\end{eqnarray}}
\newcommand{\cont}{\nonumber\eea\bea}
\newcommand{\cl}[1]{\begin{center} {#1} \end{center}}
\newcommand{\ea}{\end{array}}
%\newcommand{\arr}{\bea}

% -------------- MATH def -------------------------------
\newcommand{\ab}{{\alpha\beta}}
\newcommand{\cd}{{\gamma\delta}}
\newcommand{\dc}{{\delta\gamma}}
\newcommand{\ac}{{\alpha\gamma}}
\newcommand{\bd}{{\beta\delta}}
\newcommand{\abc}{{\alpha\beta\gamma}}
\newcommand{\eps}{{\epsilon}}
\newcommand{\lam}{{\lambda}}
\newcommand{\mn}{{\mu\nu}}
\newcommand{\mpnp}{{\mu'\nu'}}
\newcommand{\Amuu}{{A_{\mu}}}
\newcommand{\Amuo}{{A^{\mu}}}
\newcommand{\Vmuu}{{V_{\mu}}}
\newcommand{\Vmuo}{{V^{\mu}}}
\newcommand{\Anuu}{{A_{\nu}}}
\newcommand{\Anuo}{{A^{\nu}}}
\newcommand{\Vnuu}{{V_{\nu}}}
\newcommand{\Vnuo}{{V^{\nu}}}
\newcommand{\Fmnu}{{F_{\mu\nu}}}
\newcommand{\Fmno}{{F^{\mu\nu}}}

\newcommand{\abcd}{{\alpha\beta\gamma\delta}}

% Boldmath definitions

\newcommand{\bsigma}{\mbox{\boldmath $\sigma$}}
\newcommand{\beps}{\mbox{\boldmath $\varepsilon$}}
\newcommand{\btau}{\mbox{\boldmath $\tau$}}
\newcommand{\brho}{\mbox{\boldmath $\rho$}}
\newcommand{\bpipi}{\mbox{\boldmath $\pi\pi$}} 
\newcommand{\bss}{\bsigma\!\cdot\!\bsigma}
\newcommand{\btt}{\btau\!\cdot\!\btau}
\newcommand{\bnabla}{\mbox{\boldmath $\nabla$}}
\newcommand{\bphi}{\mbox{\boldmath $\tau$}}
\newcommand{\bvarphi}{\mbox{\boldmath $\rho$}}
\newcommand{\bE}{\mbox{\boldmath $E$}}
\newcommand{\bDelta}{\mbox{\boldmath $\Delta$}}
\newcommand{\bGamma}{\mbox{\boldmath $\Gamma$}}
\newcommand{\bpsi}{\mbox{\boldmath $\psi$}}
\newcommand{\bPsi}{\mbox{\boldmath $\Psi$}}
\newcommand{\bPhi}{\mbox{\boldmath $\Phi$}}
\newcommand{\bnab}{\mbox{\boldmath $\nabla$}}
\newcommand{\bpi}{\mbox{\boldmath $\pi$}}
\newcommand{\btheta}{\mbox{\boldmath $\theta$}}
\newcommand{\bkappa}{\mbox{\boldmath $\kappa$}}
\newcommand{\bgamma}{\mbox{\boldmath $\gamma$}}

\newcommand{\bp}{\mbox{\boldmath $p$}}
\newcommand{\ba}{\mbox{\boldmath $a$}}
\newcommand{\bq}{\mbox{$\vec{q}_{\perp}$}}
\newcommand{\br}{\mbox{\boldmath $r$}}
\newcommand{\bs}{\mbox{\boldmath $s$}}
\newcommand{\bk}{\mbox{$\vec{k}_{\perp}$}}
\newcommand{\bl}{\mbox{\boldmath $l$}}
\newcommand{\bb}{\mbox{\boldmath $b$}}
\newcommand{\be}{\mbox{$\vec{e}_{\perp}$}}
\newcommand{\bP}{\mbox{\boldmath $P$}}
\newcommand{\bV}{\mbox{\boldmath $V$}}
\newcommand{\bI}{\mbox{\boldmath $I$}}
\newcommand{\bJ}{\mbox{\boldmath $J$}}

\newcommand{\bT}{{\bf T}}
\newcommand{\fph}{${\cal F}$}
\newcommand{\aph}{${\cal A}$}
\newcommand{\dph}{${\cal D}$}
\newcommand{\fpi}{f_\pi}
\newcommand{\mpi}{m_\pi}
\newcommand{\Tr}{{\mbox{\rm Tr}}}
\def\Qb{\overline{Q}}
\newcommand{\delu}{\partial_{\mu}}
\newcommand{\delo}{\partial^{\mu}}
% ------------------ arrow mod ---------------------
\newcommand{\up}{\!\uparrow}
\newcommand{\upup}{\uparrow\uparrow}
\newcommand{\updo}{\uparrow\downarrow}
\newcommand{\uu}{$\uparrow\uparrow$}
\newcommand{\ud}{$\uparrow\downarrow$}
\newcommand{\auu}{$a^{\uparrow\uparrow}$}
\newcommand{\aud}{$a^{\uparrow\downarrow}$}
\newcommand{\pu}{p\!\uparrow}
% ------------------------------------------------------
\newcommand{\qp}{quasiparticle}
\newcommand{\sa}{scattering amplitude}
\newcommand{\ph}{particle-hole}
\newcommand{\qcd}{{\it QCD}}
\newcommand{\integ}{\int\!d}
\newcommand{\ie}{{\sl i.e.~}}
\newcommand{\etal}{{\sl et al.~}}
\newcommand{\etc}{{\sl etc.~}}
\newcommand{\rhs}{{\sl rhs~}}
\newcommand{\lhs}{{\sl lhs~}}
\newcommand{\eg}{{\sl e.g.~}}
\newcommand{\ef}{\epsilon_F}
\newcommand{\sigt}{d^2\sigma/d\Omega dE}
\newcommand{\sige}{{d^2\sigma\over d\Omega dE}}
% ----------------------- ------------------------------
\newcommand{\rpaeq}{\beq
\left ( \begin{array}{cc}
A&B\\
-B^*&-A^*\end{array}\right )
\left ( \begin{array}{c}
X^{(\kappa})\\Y^{(\kappa)}\end{array}\right )=E_\kappa
\left ( \begin{array}{c}
X^{(\kappa})\\Y^{(\kappa)}\end{array}\right )
\eeq}

\newcommand{\ket}[1]{{#1} \rangle}
\newcommand{\bra}[1]{\langle {#1} }

\newcommand{\Bigket}[1]{{#1} \Big\rangle}
\newcommand{\Bigbra}[1]{\Big\langle {#1} }

\newcommand{\ave}[1]{\langle {#1} \rangle}
\newcommand{\Bigave}[1]{\left\langle {#1} \right\rangle}
\newcommand{\half}{{\frac{1}{2}}}

\newcommand{\singlespace}{
    \renewcommand{\baselinestretch}{1}\large\normalsize}
\newcommand{\doublespace}{
    \renewcommand{\baselinestretch}{1.6}\large\normalsize}
\newcommand{\bftau}{\mbox{\boldmath $\tau$}}
\newcommand{\bfalpha}{\mbox{\boldmath $\alpha$}}
\newcommand{\bfgamma}{\mbox{\boldmath $\gamma$}}
\newcommand{\bfxi}{\mbox{\boldmath $\xi$}}
\newcommand{\bfbeta}{\mbox{\boldmath $\beta$}}
\newcommand{\bfeta}{\mbox{\boldmath $\eta$}}
\newcommand{\bfpi}{\mbox{\boldmath $\pi$}}
\newcommand{\bfphi}{\mbox{\boldmath $\phi$}}
\newcommand{\bfR}{\mbox{\boldmath ${\cal R}$}}
\newcommand{\bfL}{\mbox{\boldmath ${\cal L}$}}
\newcommand{\bfM}{\mbox{\boldmath ${\cal M}$}}
\def\dblint{\mathop{\rlap{\hbox{$\displaystyle\!\int\!\!\!\!\!\int$}}
    \hbox{$\bigcirc$}}}
\def\ut#1{$\underline{\smash{\vphantom{y}\hbox{#1}}}$}

\def\UNITY{{\bf 1\! |}}
\def\Pom{{\bf I\!P}}
\def\lsim{\mathrel{\rlap{\lower4pt\hbox{\hskip1pt$\sim$}}
    \raise1pt\hbox{$<$}}}         %less than or approx. symbol
\def\gsim{\mathrel{\rlap{\lower4pt\hbox{\hskip1pt$\sim$}}
    \raise1pt\hbox{$>$}}}         %greater than or approx. symbol

\newcommand\scalemath[2]{\scalebox{#1}{\mbox{\ensuremath{\displaystyle #2}}}}

\newcommand{\RP}[1]{{\blue RP: #1}}
\newcommand{\WS}[1]{{\red WS: #1}}

%%%%%%%%%%%%%%%%%%%%%%%%%%%%%%%%%%%%%%%%%%%%%%%%%%%%%%
\title{ 
%$\gamma^* \gamma$ transition form factors of $2^{++}$ quarkonia
$\chi_{c2}$ tensor meson transition form factors in the light front approach
}
%%%%%%%%%%%%%%%%%%%%%%%%%%%%%%%%%%%%%%%%%%%%%%%%%%%%%%

\author{Izabela Babiarz}
\email{izabela.babiarz@ifj.edu.pl}
\affiliation{Institute of Nuclear Physics, Polish Academy of Sciences, 
ul. Radzikowskiego 152, PL-31-342 Krak{\'o}w, Poland}

\author{Roman Pasechnik}
\email{roman.pasechnik@fysik.lu.se}
\affiliation{Department of Physics,
Lund University, SE-223 62 Lund, Sweden}

\author{Wolfgang Sch\"afer}%
\email{wolfgang.schafer@ifj.edu.pl}
\affiliation{Institute of Nuclear
Physics, Polish Academy of Sciences, ul. Radzikowskiego 152, PL-31-342 
Krak{\'o}w, Poland}

\author{Antoni Szczurek}
\email{antoni.szczurek@ifj.edu.pl}
\affiliation{Institute of Nuclear
Physics, Polish Academy of Sciences, ul. Radzikowskiego 152, PL-31-342 
Krak{\'o}w, Poland}
\affiliation{College of Mathematics and Natural Sciences,
University of Rzesz\'ow, ul. Pigonia 1, PL-35-310 Rzesz\'ow, Poland\vspace{5mm}}

\begin{abstract}
We continue our  work on the light-front formulation 
of quarkonium $\gamma^* \gamma$ transition form factors, extending the formalism to $J^{PC} = 2^{++}$ tensor meson states. We present an analysis of $\gamma^* \gamma \to \chi_{c2}$ transition amplitude and the pertinent helicity form factors. Our relativistic formalism is based on the light-front quark-antiquark wave function of the quarkonium. We calculate the two-photon decay width as well as three independent $\gamma^* \gamma$ transition form factors for $J_z = 0,1,2$ as a function of photon virtuality $Q^2$. We compare our results for the two-photon decay width to the recently measured ones by the Belle and BES III collaborations. Even when including relativistic corrections, a very small $\Gamma(\lambda = 0)/\Gamma(\lambda = 2)\sim10^{-3}$ ratio is found which is beyond present experimental precision. We also present the form factors as a function of photon virtuality and compare them to the sparse experimental data on the so-called off-shell width. The formalism presented here can be used for other $2^{++}$ mesons, excited charmonia or bottomonia or even light $q \bar q$-mesons.
\end{abstract}

\maketitle

\section{Introduction}
The production of $C$-even quarkonia in $\gamma^* \gamma$ fusion processes keeps providing us with important information on their structure \cite{Belle:2017egg,Belle:2017xsz,Belle:2022exn,Belle:2023yvd,BESIII:2012uyb,BESIII:2017rpg,CLEO:PhysRevD.78.09150,Beloborodov:2022byx}. While untagged $e^+ e^-$ cross sections give access to the decay width of quarkonia into $\gamma \gamma$ pair, in single tagged collisions, transition form factors involving one virtual and one real photon can be measured.

Here, we continue our work on the light-front formulation of $\gamma^* \gamma^*\to \chi$ transition form factors for a given meson state $\chi$. We have already presented the formalism for computing the $\gamma^* \gamma^*$ transition amplitudes to $0^\pm, 1^+$ charmonia using light-front $c \bar c$ wave functions (LFWFs) \cite{Babiarz:2019sfa,Babiarz:2020jkh,Babiarz:2022xxm,Babiarz:2023ebe}. We adopt two different approaches to the LFWFs. In the first one, they are obtained from the radial wavefunctions in a potential model, supplemented by a Melosh-transform of the relevant spin-orbit structure. The second is based on direct solutions of the bound-state problem formulated on the light-front (LF). Here, convenient tables of the wave function from the Basis Light Front Quantization (BLFQ) approach of Refs.~\cite{Li:2021ejv, Li:PhysRevD.96.016022} are available in the literature \cite{Li:PRD}.

In this work, we wish to extend the formalism to the $\gamma^* \gamma^* \to \chi_{c2}$ transition amplitude based on the quarkonium LFWF. For this purpose, we focus on the form factors describing such a coupling for one real and one spacelike virtual photon as a function of the photon virtuality. Only very sparse data are available on this process at the moment, while in principle experiments such as Belle can provide such data in the future. Recently, the Belle collaboration has measured the radiative decay width \cite{Belle:2022exn}, where they select two quasi-real photon collisions in no-tag mode.

The paper is organised as follows. First, we discuss how the current transition matrix elements for one virtual photon are related to the LFWF. In the next section, we derive the corresponding form factors. We present numerical results for the transition form factors, also in the non-relativistic approximation using $c\bar c$ wave function obtained by solving the Schr\"odinger equation. We then compare our results for the radiative decay width to available measurements.

%%%---
\section{Transition matrix elements for one real and one virtual photon}
\label{sec:Transition_ME}
%%%---

\begin{figure}[h!]
    \centering
    \includegraphics[width = 0.5\textwidth]{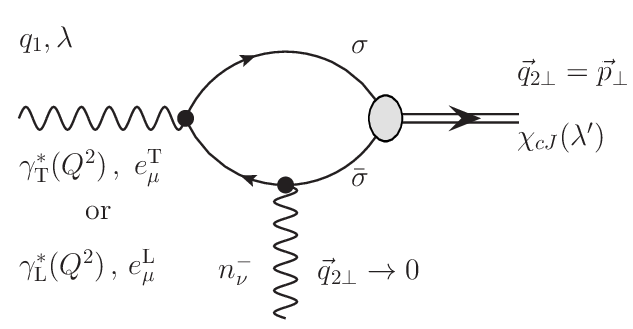}
    \caption{An example diagram for one virtual photon transition, with $q_{1} = (q_1^+, q_1^- = -\frac{Q^2}{2q_1^+}, \bq_1 =0)$, $q_{2} = (q_2^+=0,q_2^-  = P^- - q_1^-,\vec{q}_{2\perp})$.}
    \label{fig:diag}
\end{figure}

As in our recent work on the $1^{++}$ states \cite{Babiarz:2023ebe}, we start with formulating the $\gamma^* \gamma \to 2^{++}$ process in a Drell-Yan frame, in which one of the photons carries vanishing light-front plus momentum (for notation, see Fig.~\ref{fig:diag}). The relevant four-momentum transfer satisfies $q_2^2 = - {\vec{q}_{2\perp}}^{\,2}$, and we approach the on-shell limit for this photon by letting its transverse momentum go to zero $\vec q_{2\perp} \to 0$ for a meson in an external electromagnetic field. The process can therefore be viewed as a dissociation of an incoming virtual photon in an external electromagnetic field. We chose the polarization vector of the latter such that we project on the light-front plus component of the current. This choice of the frame and the current is the preferred one for the evaluation of electroweak transition currents of hadrons, as it is free from parton-number changing transitions, and instantaneous (in LF time) fermion exchanges \cite{Brodsky:1998hn}.
%%%%
%%%%%%%%%%
The pertinent helicity amplitudes are then related to matrix elements of the LF-plus component of the current as
%%%%%
\begin{multline}    
   {\cal M}(\lambda \to \lambda') \equiv \bra{\chi_{cJ}}(\lambda')| J_+(0) |\ket{\gamma^*_{\rm T,L}(Q^2)}\\
    = 2 q^{+}_{1} \, \sqrt{N_c}\,e^2 e^2_f \int \frac{dz d^2\bk} {z(1-z) 16 \pi^3} 
    \sum_{\sigma, \bar \sigma} \Psi^{\lambda'\,*}_{\sigma \bar \sigma} (z, \bk)
    (\vec{q}_{2\perp} \cdot \nabla_{\bk}) \Psi^{\gamma_{\rm T,L}}_{\sigma \bar \sigma}(z, \bk, Q^2)\, .
    \label{eq:J_plus_WF}
\end{multline}    
%%%
Here, $\sigma (\bar \sigma)$ denotes the (anti)quark polarization, and in what follows we will represent the helicities $\pm \sigma/2$ and $\pm \bar \sigma /2$ by $\uparrow$ and $\downarrow$. The fine structure constant is $\alpha_{\rm em} = e^2/(4\pi)$, $e_f$ is the electric charge of quark with flavour $f$ and with mass $m_f$. The derivative operator $(\vec{q}_{2\perp} \cdot \nabla_{\bk})$ is acting on the LFWF of the transverse $\Psi^{\gamma_T}_{\sigma\bar \sigma}$ or longitudinal $\Psi^{\gamma_L}_{\sigma\bar \sigma}$ photon, we do not explicitly display the photon polarization $\lambda$.

The explicit form of the photon LFWFs reads (see e.g.~Ref.~\cite{Kovchegov:2012mbw})
%%%
\begin{eqnarray}
    \Psi^{\gamma_T}_{\sigma \bar \sigma}(z, \bk, Q^2) &=& \sqrt{z(1-z)}\,
    \frac{
    \delta_{\sigma,-{\bar \sigma}}\, (\be \cdot \bk) \, \Big(  2(1-z)\delta_{\bar\sigma, \lambda} - 2z\delta_{\sigma,\lambda} \Big) + \delta_{\sigma \bar\sigma}\delta_{\sigma\lambda} \sqrt{2}m_f   }
    {  \bk^2 + m_f^2 +z(1-z)Q^2}\, ,\\
    \Psi^{\gamma_L}_{\sigma \bar \sigma}(z, \bk, Q^2) &=& \Big(\sqrt{z(1-z)}\Big)^3 \,
    \frac{2Q\,\delta_{\sigma,\, - \bar \sigma}}{\bk^2 +m^2_f + z(1-z)Q^2} \,, 
    \label{eq:photon_WF}
\end{eqnarray}
where $m_f$ is (anti)quark mass, and $z = k^+/q^+$ is the light front momentum fraction of photon carried by the quark and $(1-z)$ by the antiquark. Here, we defined $\varepsilon^2 = m^2_f + z(1-z) Q^2$.
%%%%%
Inserting the photon LFWFs into Eq.(\ref{eq:J_plus_WF}), we obtain for the transverse photon with helicity $\lambda = +1$:
%%%
\begin{multline}
\bra{\chi_{cJ}(\lambda')}| J_+(0) |\ket{\gamma_T^*(+1,Q^2)} = - 2 q_{1+} \sqrt{N_c} e^2 e_f^2 \int \frac{dz d^2 \bk}{\sqrt{z(1-z)} 16 \pi^3} \Big\{ \Psi^{\lambda' *}_{\uparrow \uparrow}(z,\bk) \frac{ 2 \sqrt{2} m_f (\vec{q}_{2\perp} \cdot \bk)}{[\bk^2 + \varepsilon^2]^2} \\
+ \Big( 2z \Psi^{\lambda' *}_{\uparrow \downarrow}(z,\bk) - 2 (1-z)  \Psi^{\lambda' *}_{\downarrow \uparrow}(z,\bk) \Big) \Big( \frac{\be(+) \cdot \vec{q}_{2\perp}}{\bk^2 + \varepsilon^2} - \frac{2 (\vec{q}_{2\perp} \cdot \bk)(\be(+) \cdot \bk)}{[\bk^2 + \varepsilon^2]^2}\Big) \Big\}\, , 
\end{multline}
%%%%
and for the incoming longitudinal photon
%%%
\begin{multline}
    \bra{\chi_{cJ}(\lambda')}|J_+(0)|\ket{\gamma^*_L(Q^2)} =- 2 q_{1+} \sqrt{N_c} e^2 e_f^2 \, 2 Q \,\int \frac{dz d^2 \bk}{\sqrt{z(1-z)} 16 \pi^3} z(1-z) \frac{2 \vec{q}_{2\perp} \cdot \bk}{[\bk^2 + \varepsilon^2]^2} \\
    \Big(\Psi^{\lambda' *}_{\uparrow \downarrow}(z,\bk) + \Psi^{\lambda' *}_{\downarrow \uparrow}(z,\bk) \Big)\, .
\end{multline}
%%%
Now we wish to perform the azimuthal angle integration.
To this end, we note that
%%%%
\begin{eqnarray}
    \be(+) \cdot \vec{q}_{2\perp} &=& - \frac{1}{\sqrt{2}} ( q_{2x} + i q_{2y} ) = - \frac{q_{2\perp}}{\sqrt{2}} \, e^{i \varphi_q} \, , \, \be(+) \cdot \bk =  - \frac{k_\perp}{\sqrt{2}} \, e^{i \varphi} \nonumber \\
    \vec{q}_{2\perp} \cdot \bk &=& q_{2\perp} k_\perp \cos(\varphi_q - \varphi) =  q_{2\perp} k_\perp \half \Big( e^{i \varphi_q} e^{-i \varphi} + e^{-i \varphi_q} e^{i \varphi} \Big) \, . 
\end{eqnarray}
%%%%%
In addition to these angular dependencies, also the LFWF depends on the azimuthal angle $\varphi$ of $\vec k_\perp$. Indeed, our LFWFs
\begin{eqnarray}
    \hat J_z = \hat S_z + \hat L_z \,,
\end{eqnarray}
%%%%%
which acts on the WFs as
%%%%%
\begin{eqnarray}
 \hat J_z \Psi_{\sigma \bar \sigma}^{\lambda'} (z,\vec k_\perp)  = \lambda' \, \Psi^{\lambda'}_{\sigma \bar \sigma}(z,\vec k_\perp) = \Big( \frac{\sigma + \bar \sigma}{2} - i \frac{\partial}{\partial \varphi} \Big) \Psi^{\lambda'}_{\sigma \bar \sigma}(z,\vec k_\perp) \, , 
\end{eqnarray}
%%%%%
so that we can isolate the $\varphi$ dependence as
%%%%%
\begin{eqnarray}
   \Psi^{\lambda'}_{\sigma \bar \sigma}(z,\vec k_\perp) = \tilde \psi^{\lambda'}_{\sigma \bar \sigma}(z, k_\perp) \, e^{i L_z \varphi} \, , \quad  {\rm with} \quad L_z = \lambda' - S_z \, .
\end{eqnarray}
%%%%%
As a result,
%%%%%
\begin{eqnarray}
\Psi^{\lambda'}_{\uparrow \uparrow}(z, \bk) = \tilde \psi^{\lambda'}_{\uparrow \uparrow}(z,k_\perp) \, e^{i (\lambda' - 1) \varphi} \, , \, \Psi^{\lambda'}_{\uparrow \downarrow}(z, \bk) = \tilde \psi^{\lambda'}_{\uparrow \downarrow}(z,k_\perp) \, e^{i \lambda' \varphi} \, , \, \Psi^{\lambda'}_{\downarrow \uparrow}(z, \bk) = \tilde \psi^{\lambda'}_{\downarrow \uparrow}(z,k_\perp) \, e^{i \lambda' \varphi} \, . \nonumber 
\end{eqnarray}
%%%%%
We can now straightforwardly perform the angular integration:
%%%%%
\begin{multline}
\bra{\chi_{cJ}(\lambda')}| J_+(0) |\ket{\gamma_T^*(+1,Q^2)} = -  2 q_{1+} \sqrt{2 N_c} e^2 e_f^2 \, \Big\{
(q_{2x} + i q_{2y})  \delta_{\lambda',0}
\int \frac{dz \, k_\perp dk_\perp}{\sqrt{z(1-z)} 8 \pi^2} 
  \frac{1}{[k_\perp^2 + \varepsilon^2]^2} \\ 
  \times \Big[ m_f k_\perp  \tilde \psi^{\lambda' }_{\uparrow \uparrow}(z,k_\perp)
  - \varepsilon^2 \Big( z \tilde \psi^{\lambda' }_{\uparrow \downarrow}(z,k_\perp) - (1-z)  \tilde \psi^{\lambda'}_{\downarrow \uparrow}(z,k_\perp) \Big) \Big] \\
  + (q_{2x} - i q_{2y})  \delta_{\lambda',2}\int \frac{dz \, k_\perp dk_\perp}{\sqrt{z(1-z)} 8 \pi^2} 
  \frac{1}{[k_\perp^2 + \varepsilon^2]^2} \\ 
  \times \Big[ m_f k_\perp  \tilde \psi^{\lambda' }_{\uparrow \uparrow}(z,k_\perp)
  + k_\perp^2 \Big( z \tilde \psi^{\lambda' }_{\uparrow \downarrow}(z,k_\perp) - (1-z)  \tilde \psi^{\lambda'}_{\downarrow \uparrow}(z,k_\perp) \Big) \Big] \Big\}
  \, . 
\end{multline}
%%%%%
In the same manner, we obtain for the transitions of the longitudinal photon:
%%%%%
\begin{multline}
   \bra{\chi_{cJ}(\lambda')}|J_+(0)|\ket{\gamma^*_L(Q^2)} 
   = - 2 q_{1+} \sqrt{N_c} e^2 e_f^2 \, 2 Q  \Big( (q_{2x} + i q_{2y}) \delta_{\lambda',-1} + (q_{2x} - i q_{2y}) \delta_{\lambda',+1} \Big) \\
   \times \int \frac{dz k_\perp dk_\perp}{\sqrt{z(1-z)} 8 \pi^2} \frac{z(1-z)k_\perp}{[k^2_\perp + \varepsilon^2]^2} \Big(\tilde \psi^{\lambda' }_{\uparrow \downarrow}(z,k_\perp) + \tilde \psi^{\lambda' }_{\downarrow \uparrow}(z,k_\perp) \Big)\, .
\end{multline}

The procedure for obtaining the LFWFs for the spin-two state is described in Appendix~\ref{app:LFWF_melosh}.

%%%%----------------------------------------------------
\section{Form factors $\gamma \gamma^* \to 2^{++}$}
%%%%----------------------------------------------------

Now, we wish to express our results for the transition amplitudes
in the Drell-Yan frame through the invariant transition form factors commonly used in the literature. For definiteness, here we use the form factors introduced in Ref.~\cite{Poppe:1986dq}, while for different conventions, see e.g.~Ref.~\cite{Pascalutsa:2012pr, Schuler:1997yw}.

We start from the parametrization of the covariant amplitude for the process $\gamma^*(q_1) \gamma(q_2) \to 2^{++}$\footnote{We have simplified the notation in Ref.~\cite{Poppe:1986dq} by introducing
\begin{eqnarray}
    F_{\rm LT}(Q^2) = (q^2_2 - q^2_1)F'_{\rm TL} - F_{\rm TL}\, . \nonumber
\end{eqnarray}
}
:
%%%%
\begin{eqnarray}
\frac{1}{4 \pi \alpha_{em}}{\cal M}_{\mu \nu \alpha \beta} &=& \delta_{\mu \nu}^\perp (q_2 - q_1)_\alpha (q_2 - q_1)_\beta \, F_{\rm TT,0}(Q^2) + \delta^\perp_{\mu \alpha} \delta^\perp_{\nu \beta} \, F_{\rm TT,2}(Q^2) \nonumber \\
&&+  \Big( q_{1\mu} -\frac{q^2_1}{q_1 \cdot q_2} q_{2\mu} \Big) \delta^\perp_{\nu \alpha} (q_2 - q_1)_\beta \, F_{\rm LT}(Q^2)  \, ,
\end{eqnarray}
%%%%%
where 
%%%%%
\begin{eqnarray}
\delta^{\perp}_{\mu \nu} &=& g_{\mu \nu } -\frac{1}{(q_1 \cdot q_2)^2} \Big( (q_1 \cdot q_2) (q_{2\mu}q_{1\nu} + q_{1\mu}q_{2\nu}) - q^2_1q_{2\mu}q_{2\nu} \Big)\,.
\end{eqnarray}
%%%%
Here, the four momenta of photons satisfy $q_1^2 = -Q^2,\, Q^2 \geq 0$, and $q_2^2 = 0$.

We now match the form factors defined above to the transition amplitudes calculated in the LF formalism by expressing them as
%%%%
\begin{eqnarray}
{\cal M} (\lambda \to \lambda') =
e_\mu(\lambda) n^-_\nu  {\cal M} ^{\mu \nu \alpha \beta} \, E^*_{\alpha \beta}(\lambda') \, .
\end{eqnarray}
%%%%
Introducing the light-like vectors $n^\pm_\mu$, which full fill conditions $n^+ \cdot n^+ = n^- \cdot n^- = 0$ and $n^+ \cdot n^- = 1$, we write the photon momentum as
%%%%
\begin{eqnarray}
  q_{1\mu} = q^+_1 n^+_\mu - \frac{Q^2}{2 q^+_1} n^-_\mu \, . 
\end{eqnarray}
%%%%
Further, we define the polarization of the incoming photon and outgoing meson in the LF notation, for the photon:
\begin{eqnarray}
     e_{\mu}(0) &=& \frac{1}{Q}q_{1, \mu} + \frac{Q}{q^+_1}n^-_{\mu} \,, \quad  e_{\mu}(\lambda) = e^\perp_{\mu}(\lambda)\, ,
\end{eqnarray}
and for the tensor meson:
\begin{eqnarray}    
    E^{\alpha \beta}(\pm 2) &=& E^\alpha (\pm1) E^\beta (\pm 1)\, ,
    \nonumber \\
    E^{\alpha \beta}(\pm1) &=&  \frac{1}{\sqrt{2}} \Big( E^\alpha (\pm1) E^\beta(0) + E^\alpha(0) E^\beta (\pm 1) \Big)\, , \nonumber \\
    E^{\alpha \beta}(0) &=& \frac{1}{\sqrt{6}} \Big( E^\alpha (+1) E^\beta(-1) + E^\alpha(-1) E^\beta(+1) + 2 E^\alpha(0) E^\beta(0) \Big) \, , 
\end{eqnarray}
where 
\begin{eqnarray}
        E^{\alpha}(0) &=& \frac{1}{M} P^{\alpha} - \frac{M}{P_+}n_-^{\alpha}  \,,  \quad   E^{\alpha}(\lambda) = e_\perp^{\alpha}(\lambda) - \frac{e_{\perp}(\lambda)\cdot P}{P_+} n_-^{\alpha}\,.
\end{eqnarray}
%%%%
We have denoted the four-momentum of the tensor meson as $P_\mu = q_{1\mu} + q_{2\mu}$, and notice, that $P_+ = q^+_{1}$. Above $M$ denotes the mass of the tensor meson, and $P^2 = M^2$.

Now, we can move to the transition amplitudes in the Drell-Yan frame (see Fig.~\ref{fig:diag}). 
\begin{eqnarray}
        {\cal M}(+1 \to 0) 
        &=& 2 q_1^+ \, e^2 \,(\vec{e}_{\perp}(+1) \cdot \vec{q}_{2 \perp}) \frac{2}{\sqrt{6}}\frac{M^2 + Q^2}{M^2}\, F_{\rm TT,0}(Q^2)\, , \nonumber\\
    {\cal M}(+1 \to +2) 
    &=& -2 q_1^+  e^2 \,(\vec{e}\,^*_{\perp}(+1) \cdot \vec{q}_{2 \perp}) \frac{1}{M^2 + Q^2}\, F_{\rm TT,2}(Q^2)\, ,  \nonumber\\
    {\cal M}(0\to +1)
    &=& 2 q_1^+ e^2  (\vec{e}\,^*_{\perp}(+1) \cdot \vec{q}_{2 \perp})\, \frac{Q}{\sqrt{2} M}\, F_{\rm LT}(Q^2)\, . \nonumber
\end{eqnarray}
Combining these expressions with our results for the matrix elements, we obtain the three independent transition form factors:
\begin{multline}
    F_{\rm TT,0}(Q^2) = \sqrt{6 N_c} e_f^2 \frac{M^2}{M^2 + Q^2} \int \frac{dz \, k_\perp dk_\perp}{\sqrt{z(1-z)}8 \pi^2} \, \frac{1}{[k_\perp^2 + \varepsilon^2]^2}\Bigg[ m_f k_\perp \tilde \psi^0_{\uparrow \uparrow}(z,k_\perp) \\
    - \frac{\varepsilon^2}{2} \Big( (2z -1)\Big(\tilde\psi^0_{\uparrow \downarrow}(z,k_\perp) + \tilde \psi^0_{\downarrow \uparrow}(z,k_\perp) \Big) + \Big(\tilde\psi^0_{\uparrow \downarrow}(z,k_\perp) - \tilde \psi^0_{\downarrow \uparrow}(z,k_\perp)\Big) \Big)\Bigg ]\,, 
\end{multline}
%%%%
\begin{multline}
     F_{\rm TT,2}(Q^2) = -2 \sqrt{N_c} e_f^2 (M^2 + Q^2)\int \frac{dz \, k_\perp dk_\perp}{\sqrt{z(1-z)}8 \pi^2} \, \frac{1}{[k_\perp^2 + \varepsilon^2]^2}\Bigg[ m_f k_\perp \tilde \psi^{+2}_{\uparrow \uparrow}(z,k_\perp) \\
    + \frac{k_\perp^2}{2} \Big( (2z -1)\Big(\tilde\psi^{+2}_{\uparrow \downarrow}(z,k_\perp) + \tilde\psi^{+2}_{\downarrow \uparrow}(z,k_\perp)\Big)+  \Big( \tilde\psi^{+2}_{\uparrow \downarrow}(z,k_\perp) -\tilde \psi^{+2}_{\downarrow \uparrow}(z,k_\perp) \Big )\Big)\Bigg ] \, ,
\end{multline}
\begin{multline}
    F_{\rm LT}(Q^2) =  4\sqrt{N_c} e_f^2 \, M \int \frac{dz k_\perp dk_\perp}{\sqrt{z(1-z)} 8 \pi^2} \frac{z(1-z)k_\perp}{[k^2_\perp + \varepsilon^2]^2}  \Big(\tilde \psi^{+1 }_{\uparrow \downarrow}(z,k_\perp) + \tilde \psi^{+1 }_{\downarrow \uparrow}(z,k_\perp) \Big)\,.
\end{multline}
From this representation of the transition form factors we can distinguish the ingredients related to spin-singlet  $(\tilde\psi^{\lambda'}_{\uparrow \downarrow}(z,k_\perp) -\tilde \psi^{\lambda'}_{\downarrow \uparrow}(z,k_\perp)$), as well as spin-triplet $(\tilde\psi^{\lambda'}_{\uparrow \downarrow}(z,k_\perp) + \tilde \psi^{\lambda'}_{\downarrow \uparrow}(z,k_\perp))$.
Using the formulas given in the table in Appendix A1 of Ref.~\cite{Poppe:1986dq}, these form factors can be also related to helicity amplitudes in the $\gamma^* \gamma$ c.m.~frame.
\begin{figure}[h!]
    \centering
    \includegraphics[width = 0.45\textwidth]{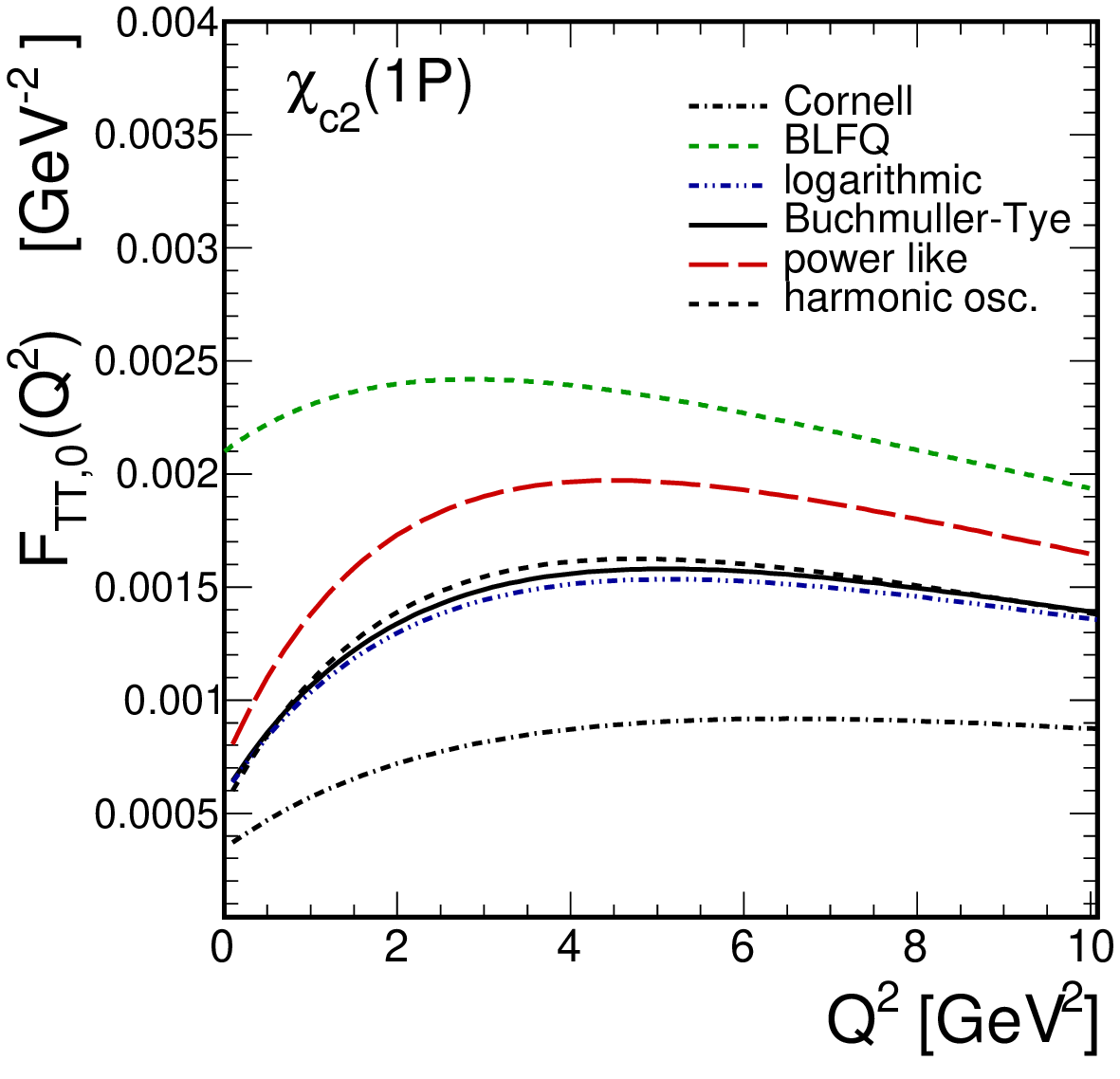}
    \includegraphics[width = 0.45\textwidth]{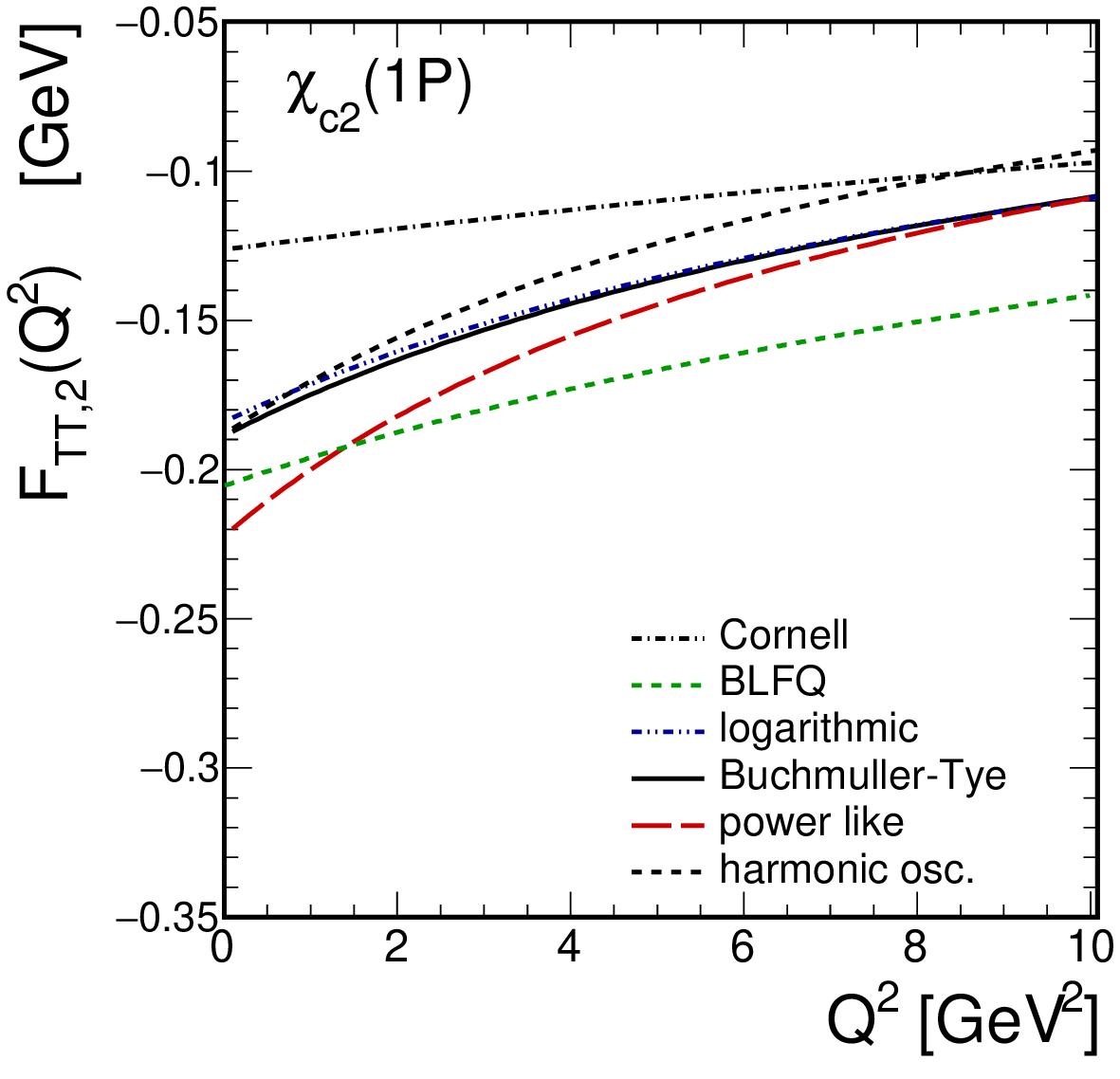}
    \includegraphics[width = 0.45\textwidth]{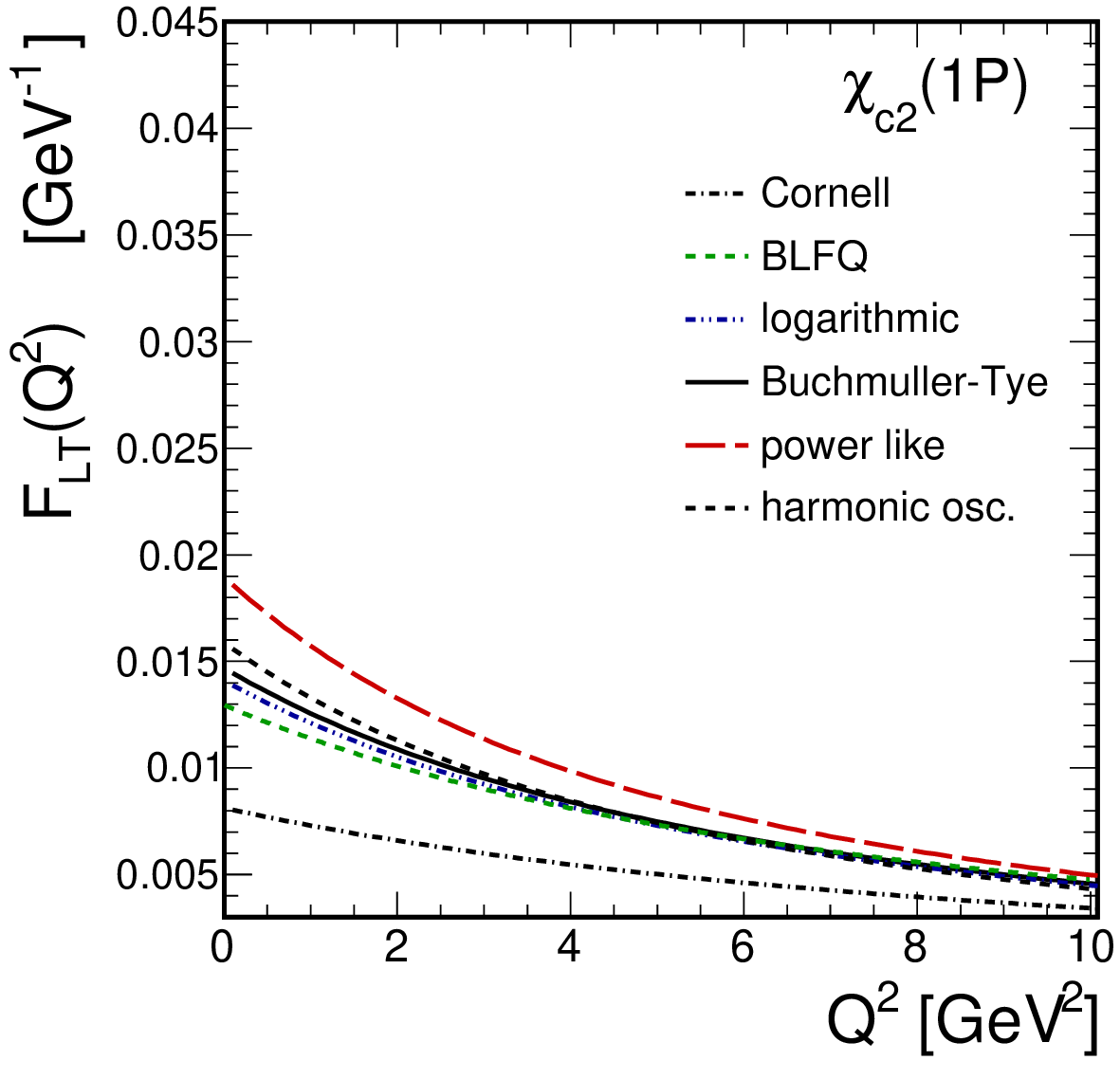}
    \caption{Three transition form factors within the LFWF approach: on the l.h.s. -- $F_{\rm TT,0}(Q^2)$, on the r.h.s. -- $F_{\rm TT,2}(Q^2)$, in the middle -- $F_{\rm LT}(Q^2)$. Here line denoted as BLFQ is a result obtained with the set of 32 expansion terms of light front wave function from the database \cite{Li:PRD}. }
    \label{fig:FF_LFWF}
\end{figure}

In Fig.~\ref{fig:FF_LFWF} we present transition form factors $F_{\rm TT,0}(Q^2)$, $F_{\rm TT,2}(Q^2)$, $F_{\rm LT}(Q^2)$ for one real and one virtual photon as a function of the photon virtuality. In the numerical calculation, we use light-front wave functions obtained
for different $c\bar c$ potentials from literature as in Ref.~\cite{Babiarz:2023ebe} or \cite{Cepila:2019skb}.
There is a relatively large spread of the results, similar to what was observed for $\gamma^*\gamma \to \chi_{c1}$ \cite{Babiarz:2023ebe}.

%%%--------------------
\subsection{NRQCD limit}
%%%--------------------

It is instructive to derive the transition form factors in the limit of nonrelativistic (NR) motion of quarks in the bound state. To reach the NR limit, we should expand the integrand around the $z=1/2$ and $\vec{k}_{\perp} = 0$, i.e.  
\begin{eqnarray}
    z = \frac{1}{2} -\xi\, , \quad  1-z = \frac{1}{2} + \xi \, , \quad \xi \to 0 \, ,
\end{eqnarray}
thus,
%%%%
\begin{eqnarray}
    z(1-z) = \frac{1}{4} - \xi^2\, , \quad  \Big( \vec{k}_{\perp}^2 + m_f^2 +z(1-z)Q^2 \Big)^2   \rightarrow (m_f^2 + Q^2/4)^2 \, .
\end{eqnarray}
%%%%
In the Melosh transform formalism described in Appendix~\ref{app:LFWF_melosh}, the LFWF can be related to the NR radial WF, $u_1(k)$. After the NR expansion, all FFs will be proportional to the integral
%%%
\begin{eqnarray}
\int_0^\infty dk \, k^2 u_1(k) \propto R'(0) \, , 
\end{eqnarray}
%%%
where $R'(0)$ is the derivative of the (spatial) radial WF at the origin, which we obtain as in Ref.~\cite{Babiarz:2020jkh}. 
As a result, the transition form factors take the form: 
\begin{eqnarray}
    F_{\rm TT,0 }(Q^2) &=& e_f^2 \, (-4)\,\sqrt{\frac{3 N_c M}{\pi}} \frac{Q^2}{(M^2 +Q^2)^3} R'(0) \, , \label{eq:FF_NRQCD_TT0}\\ 
    F_{\rm TT,2} (Q^2)  &=& e_f^2 \, 8\, \sqrt{\frac{3 N_c M}{\pi}} \frac{1}{M^2 +Q^2}\, R'(0) \, , \label{eq:FF_NRQCD_TT2}\\ 
    F_{\rm LT}(Q^2) &=& e_f^2 \, (-8)\,\sqrt{\frac{3 N_c M}{\pi}} \, \frac{1}{(M^2 + Q^2)^2}\, R'(0) \, . \label{eq:FF_NRQCD_LT}
\end{eqnarray}
Above, $M$ stands for the mass of $\chi_{c2}$ (1P), and $N_c$ is the number of colors. In the NR limit, the mass of the meson should be understood as $M = 2m_f$. These results fully agree with those obtained previously in \cite{Schuler:1997yw,Pascalutsa:2012pr}.
%%%%
\begin{figure}[h!]
    \centering
    \includegraphics[width = 0.45 \textwidth]{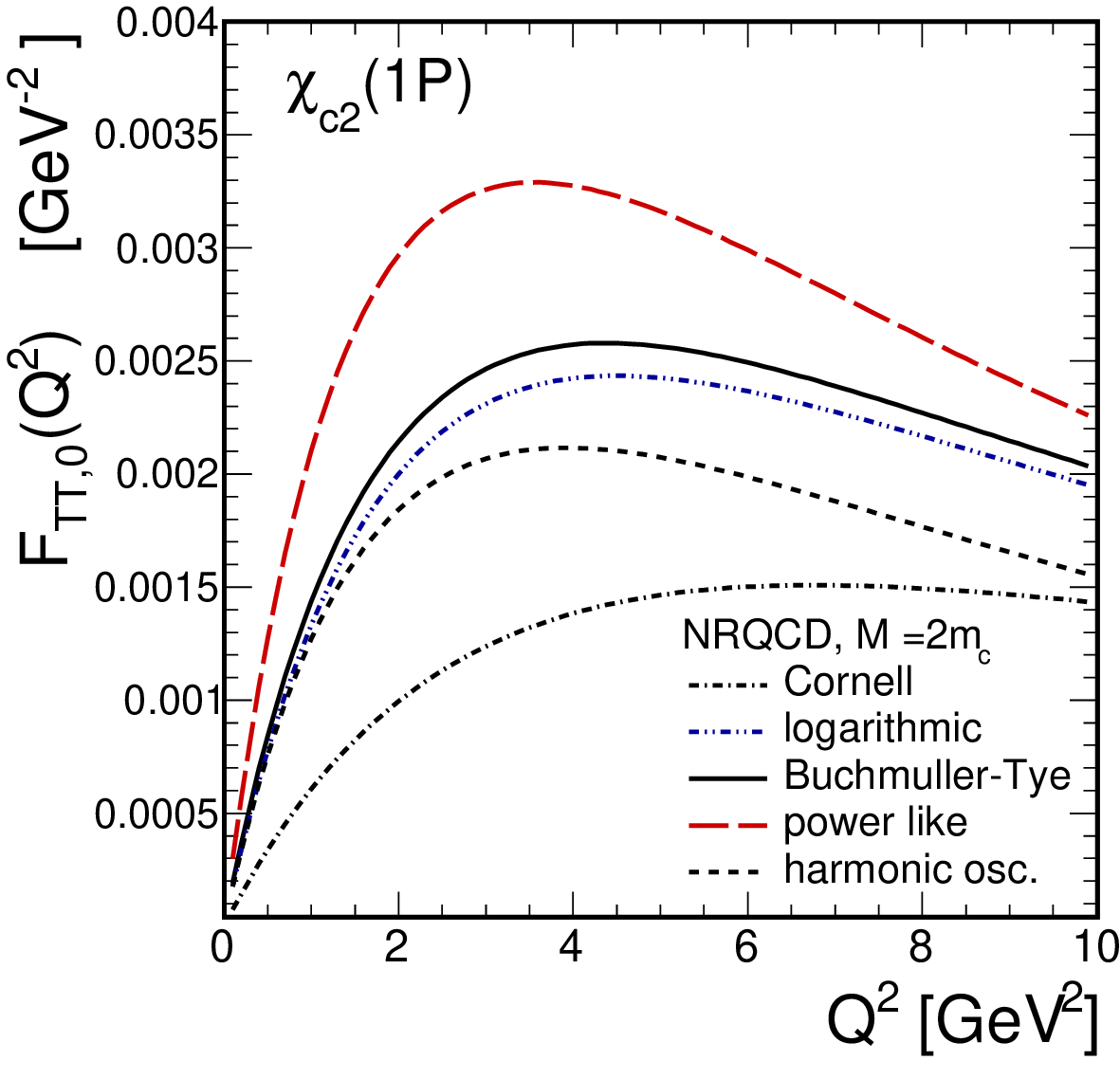}
    \includegraphics[width = 0.45 \textwidth]{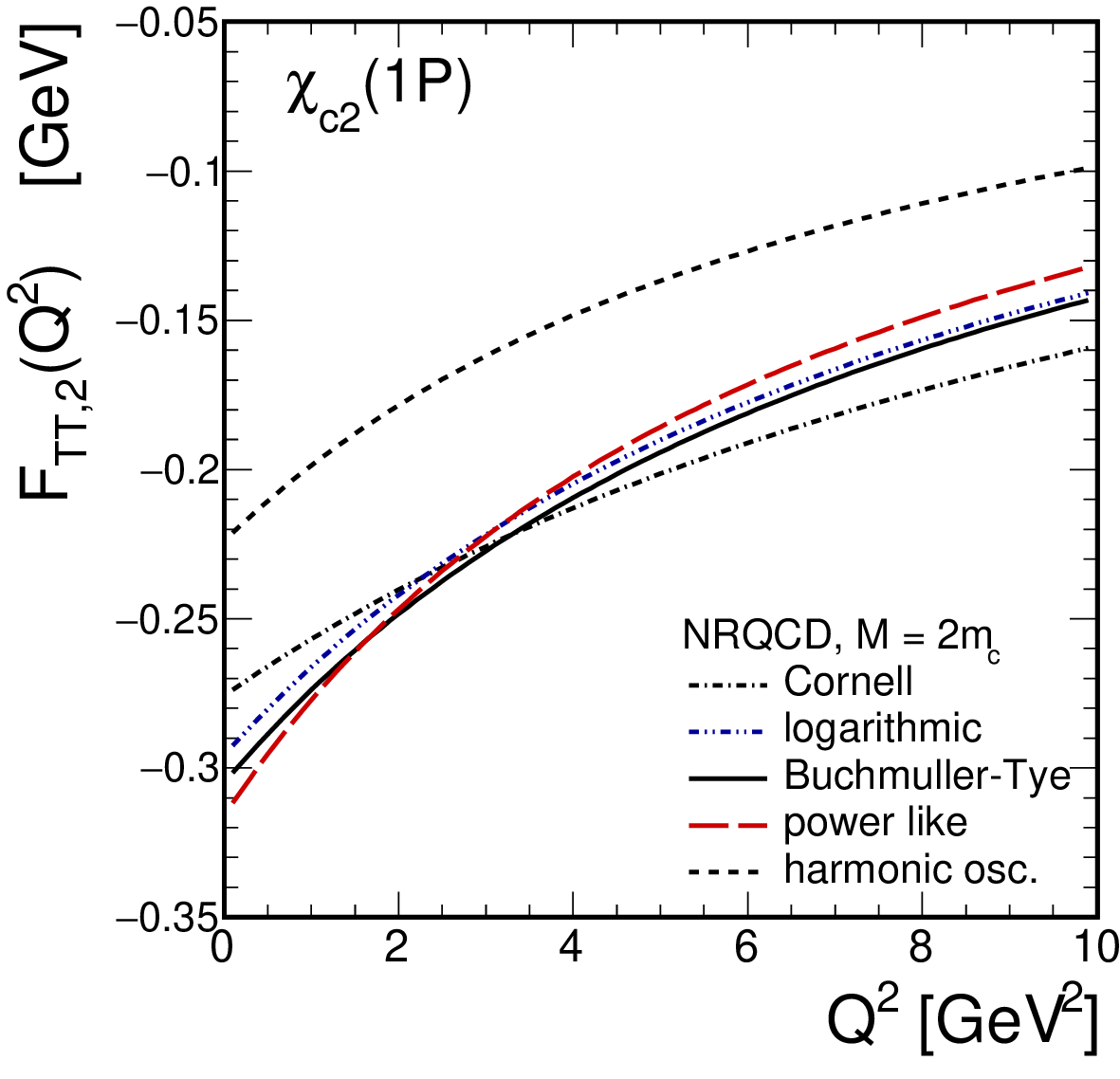}
    \includegraphics[width = 0.45 \textwidth]{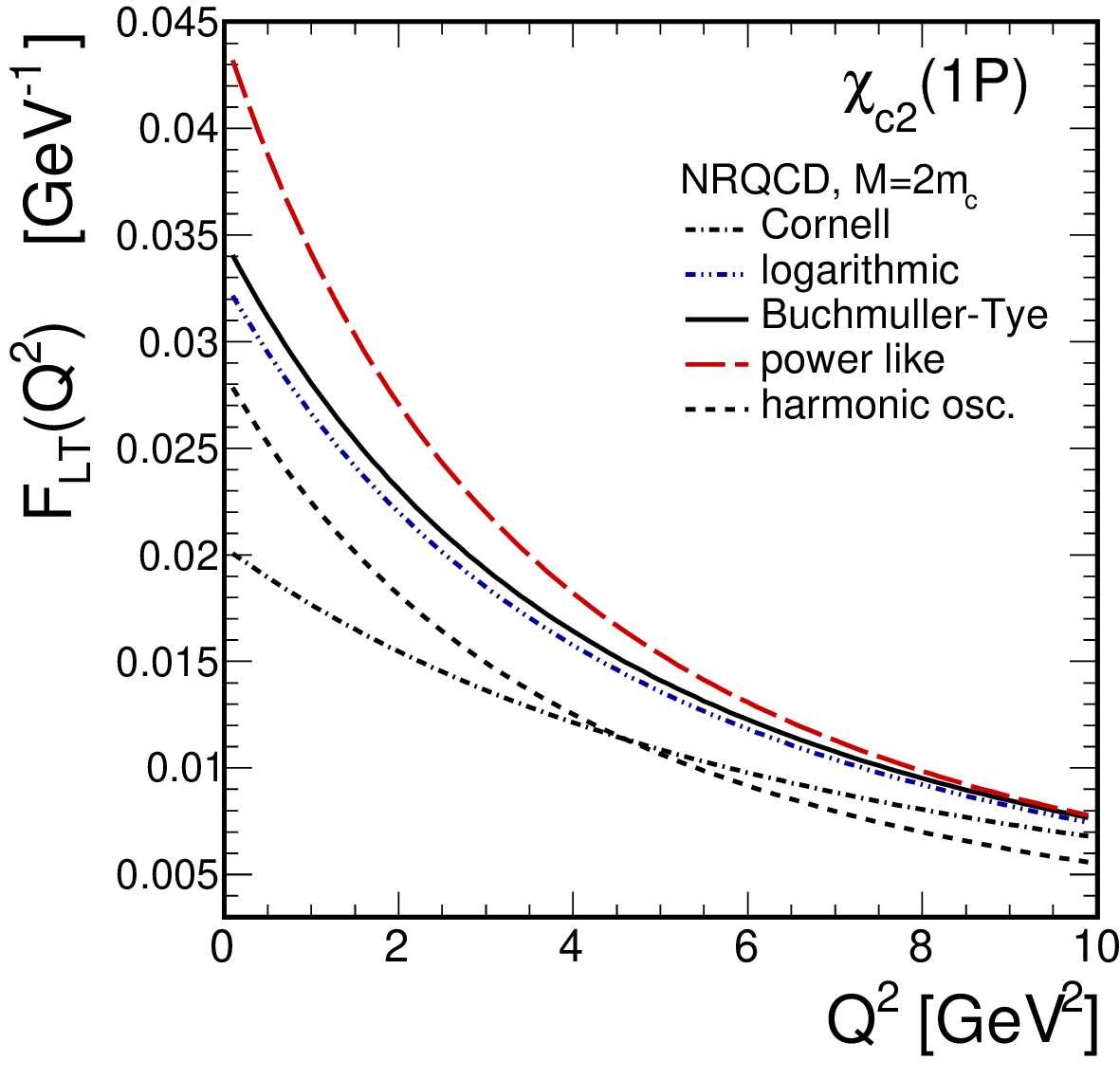}
    \caption{The three transition form factors in the non-relativistic limit (with $M=2m_f$): on the l.h.s. -- $F_{\rm TT,0}(Q^2)$, on the r.h.s. -- $F_{\rm TT,2}(Q^2)$, in the middle -- $F_{\rm LT}(Q^2)$. }
    \label{fig:FF_NRQCD_2mc}
\end{figure}
%%%

In Fig.~\ref{fig:FF_NRQCD_2mc}, we present similar results for the non-relativistic approach with $M=2m_f$, see Eqs.~(\ref{eq:FF_NRQCD_TT0}) -- (\ref{eq:FF_NRQCD_LT}). We hope that in the near future, such form factors will be extracted by the Belle collaboration. So far, only $\Gamma_{\gamma\gamma}(Q^2)$ as defined by the Belle collaboration was measured.

%%-------------------------------------------------
\section{$2^{++} \to \gamma \gamma$ decay width}
%%-------------------------------------------------

The radiative decay width is described by two contributions from $J_z = 2$ ($F_{\rm TT,2}$), and $J_z = 0$ ($F_{\rm TT,0}$): 
\begin{eqnarray}
  \Gamma_{\gamma\gamma}(\chi_{c2}) = (4\pi \alpha_{\rm em})^2 \Bigg[ \frac{|F_{\rm TT,0}(0)|^2 \cdot M_{\chi_{c2}}^3}{120\pi} + \frac{|F_{\rm TT,2} (0)|^2}{80 \pi M_{\chi_{c2}}} \Bigg] \,.
  \label{eq:Gamma_FF}
\end{eqnarray}
Therefore, we can neglect the $J_z =\pm 1$ contribution related to $F_{\rm LT}$  in no-tag mode. Nevertheless, one would expect the cross-section $\sigma_{\rm TT}(J_z = 0)$ to be considerably smaller than  $\sigma_{\rm TT}(J_z =\pm  2)$. In further calculation we take $M_{\chi_{c2}} = 3.556~{\rm GeV}$ \cite{PDG_Workman:2022ynf}. 
%%%%%%%%%%%%%%%%%%%%%%%%%%%%%%%%%%%%%%%%%%%%%%
\begin{table}[h!]
    \caption{Transition form factors for $\chi_{c2}$(1P) at the on-shell point with corresponding $c$-quark mass.}
    \centering
    \begin{tabular}{lccc|cc}
    \hline
    \hline
                     &  & \multicolumn{2}{c|}{LFWF}& \multicolumn{2}{c}{NRQCD}\\
                     &  &                     &   &  $M =M_{\chi_{c2}}$  & $M = 2m_f$\\
                     \hline
    potential type & $m_c$ & $F_{\rm TT,0}$ &$F_{\rm TT,2}$& $F_{\rm TT,2}$  & $F_{\rm TT,2}$\\
                        & [GeV]  & [GeV$^{-2}$]     & [GeV]  & [GeV] & [GeV]\\
    \hline
     Cornell            & 1.84   & $3.43\cdot10^{-4}$ & $-$0.13  & $-$0.29 & $-$0.28\\
     logarithmic        & 1.5    & $5.84\cdot10^{-4}$ & $-$0.18  &$-$0.23  &$-$0.30\\
     Buchm{\"u}ller-Tye & 1.48   & $5.91\cdot10^{-4}$ & $-$0.19  &$-$0.23  &$-$0.31\\
     power-like         & 1.334  & $7.20\cdot10^{-4}$ & $-$0.22  & $-$0.21 &$-$0.31\\
     harmonic osc. & 1.4   & $5.28\cdot10^{-4}$  & $-$0.19 & $-$0.16 &$-$0.22\\
     BLFQ  N32 \cite{Li:PRD} & 1.6   & $2.10\cdot10^{-3}$  & $-$0.21 &         &\\ 
     \hline
     \hline
    \end{tabular}
    \label{tab:FF_chic2}
\end{table}
%%%%%%%%%%%%%%%%%%%%%%%%%%%%%%%%%%%%%%%%%%%%%%
\begin{table}[h!]
    \caption{Helicity decomposition of the two-photon decay width of $\chi_{c2}$(1P).}
    \centering
    \begin{tabular}{lcccc|cc}
    \hline
    \hline
                 &\multicolumn{4}{c|}{LFWF}& \multicolumn{2}{c}{NRQCD}\\
                  &\multicolumn{4}{c|}{ }& $M= M_{\chi_{c2}}$ & $M = 2m_f$ \\    
                  \hline
    & $\Gamma_{\gamma\gamma}(\lambda = 0)$ & $\Gamma_{\gamma\gamma}(\lambda = \pm 2)$ & $\frac{\Gamma(\lambda =0 )}{\Gamma(\lambda = \pm 2)}$  &$\Gamma_{\gamma\gamma}$  &$\Gamma_{\gamma\gamma}(\lambda = \pm 2)$ & $\Gamma_{\gamma\gamma} (\lambda = \pm 2)$\\
                         &  [keV] & [keV]&  & [keV]   &  [keV]  &   [keV]  \\
    \hline
     Cornell             & $1.18\times 10^{-4}$ & 0.15  & $0.7\times10^{-3}$  &  0.15  & 0.79 & 0.69\\
     logarithmic         & $3.37\times 10^{-4}$ & 0.32  & $0.3\times10^{-3}$&   0.32  & 0.49 & 0.98\\
     Buchm{\"u}ller-Tye  & $3.36\times10^{-4}$  & 0.34  & $1.0\times10^{-3}$ &  0.34   & 0.51 & 1.052\\
     power like          & $5.18\times10^{-4} $ & 0.47  & $1.1\times10^{-3}$ &  0.47 & 0.40 & 1.25\\
     harmonic osc.       & $2.80 \times10^{-4}$ & 0.33  & $0.8\times10^{-3}$&  0.33  & 0.23 & 0.60\\
     BLFQ N8  & $5.18 \times10^{-3}$ & 0.39 & $1.3\times10^{-2}$& 0.39   &      & \\ 
     BLFQ N16 & $4.95 \times10^{-3}$ & 0.40 & $1.2\times10^{-2}$& 0.41   &      & \\ 
     BLFQ N24 & $4.67 \times10^{-3}$ & 0.39 & $1.2\times10^{-2}$& 0.40   &      & \\ 
     BLFQ N32 & $4.42 \times10^{-3}$ & 0.40 & $1.1\times10^{-2}$& 0.40   &      & \\ 
     \hline
     \hline
    \end{tabular}
    \label{tab:Decay_chic2}
\end{table}

The form factor at the on-shell point $F_{\rm TT,2}(0)$ in the non-relativistic limit leads to the following expression:
%%%
\begin{eqnarray}
    F_{\rm TT,2}(0)
    = 8 e_f^2 \sqrt{\frac{3N_c}{\pi M^3}} \, R'(0)\, .
\end{eqnarray}
Furthermore, as can be seen from Eq.(\ref{eq:FF_NRQCD_TT0}), in the NR limit we have $F_{\rm TT,0}(0) =0$, so that  we need to consider only the contribution from $J_z = 2$ for the radiative decay width:
\begin{eqnarray}
\Gamma_{\gamma\gamma}(\lambda = \pm  2) = \alpha^2_{\rm em} e^4_f \, \frac{3^2 2^6}{5 M^4}\, |R'(0)|^2 = \alpha^2_{\rm em} e^4_f \, \frac{36}{5 m_f^4}\, |R'(0)|^2 \, .
\end{eqnarray}
%%%
In Tab.~\ref{tab:FF_chic2} we show the values of transition form factors $F_{\rm TT,0}$ and $F_{\rm TT,2}$ for $Q^2 =0$.  In the fully relativistic calculation, we find that at $Q^2 = 0$ the $F_{\rm TT,0}$ does not vanish, but gives a negligibly small contribution. The corresponding widths $\Gamma_{\gamma\gamma}(\lambda = 0)$ and $\Gamma_{\gamma\gamma}(\lambda = \pm 2)$ in keV are shown in Tab.~\ref{tab:Decay_chic2}. Indeed, the decay width for $\lambda = 0$ is three orders of magnitude smaller than that for $\lambda = \pm 2$. We also show the ratios of the different helicity contributions to the width.
For the NR limit, where the $\lambda =0$ contribution vanishes, we show the result for $\lambda = \pm 2$ for two different approximations.

The BES III Collaboration measured the ratio between two-photon partial widths, for the $\chi_{c2}$ helicity $\lambda = 0$ and $\lambda=2$ \cite{BESIII:2017rpg}:
\begin{eqnarray}
   \frac{\Gamma_{\gamma \gamma} (\lambda=0)}{\Gamma_{\gamma \gamma}(\lambda = \pm 2)} = (0.0\pm0.6\pm 1.2) \times 10^{-2}\,,
\end{eqnarray}
which is a straightforward confirmation that the helicity-zero component is strongly suppressed. We predict the ratio of the order of $10^{-3}$. The BES III precision is not sufficient to measure the small ratios predicted in this work.

%%---------------------------------------------------
\section{Form factor $\gamma \gamma^* \to 0^{++}$}
%%---------------------------------------------------

We now want to compare the two-photon decay width for $0^{++}$ and $2^{++}$ states.
To make the comparison more transparent we reformulate the results of \cite{Babiarz:2020jkh} using the same setup in the Drell-Yan frame as in Sec. \ref{sec:Transition_ME}.
Now we have 
\begin{multline}    
    \bra{\chi_{c0}}| J_+(0) |\ket{\gamma^*_T(Q^2)}
    = 2 q^{+}_{1} \, \sqrt{N_c}\,e^2 e^2_f (\be(\lambda) \,\cdot \,\vec{q}_{2 \perp}) 2 \int \frac{dz k_{\perp} d k_{\perp}} {\sqrt{z(1-z)} 8 \pi^2}\\
    \times \Bigg\{ 
    \frac{m_f k_\perp \, {\tilde\psi}_{++}(z,k_{\perp})}{[\bk^2 + \varepsilon^2]^2}
    + \frac{\varepsilon^2}{[\bk^2 + \varepsilon^2]^2}\Big(-z\,{\tilde \psi}_{+-}(z,k_{\perp}) +(1-z)\,{\tilde \psi}_{-+}(z,k_{\perp})\Big)
    \Bigg\}\, .
\end{multline}
%%%
%%%
The helicity amplitude with the transverse photon polarization  $e^T_{\rm \mu} = (0,0,\be(\lambda))$ is obtained as
\begin{eqnarray}
    e_{\mu}^{\rm T} n^{-}_{\nu} \mathcal{M}^{\mu \nu}=  
    e_{\mu}^{\rm T} n^{-}_{\nu}  \Big(  g_{\mu \nu} - \frac{q_{1\nu} q_{2\mu}}{q_1 \cdot q_2} \Big) 4 \pi \alpha_{\rm em} F_{\rm TT}(Q^2) =  
 2 q_1^+ \frac{\vec{e}_{\perp}(\lambda) \cdot \vec{q}_{2\perp}}{M^2 + Q^2} \, 4 \pi \alpha_{\rm em} \,F_{\rm TT}(Q^2) \, , \nonumber \\
\end{eqnarray}
and $F_{\rm TT}(Q^2)$ is a function invariant under Lorentz transformation.
%%%%%
\begin{multline}    
    F_{\rm TT}(Q^2) = e_f^2\sqrt{N_c} 2(M_{\chi_{c0}}^2 +Q^2)  \int \frac{dz k_{\perp} d k_{\perp}} {\sqrt{z(1-z)} 8 \pi^2}
    \Bigg\{ 
    \frac{m_f k_\perp \, {\tilde\psi}_{++}(z,k_{\perp})}{[\bk^2 + \varepsilon^2]^2} \\
    - \frac{m^2_f + z(1-z)Q^2}{2[\bk^2 + \varepsilon^2]^2}\Big(
    (2z-1)({\tilde \psi}_{+-}(z,k_{\perp}) + {\tilde \psi}_{-+}(z,k_{\perp}))  
    + ( {\tilde \psi}_{+-}(z,k_{\perp}) - {\tilde \psi}_{-+}(z,k_{\perp}))
    \Big)
    \Bigg\}\, .
\end{multline}
%%%
For further use of LFWF calculated via the potential model and Melosh spin rotation transformation \cite{Babiarz:2020jkh}, we can find the relation between the so-called "radial" part of the light-front wave function $\psi(z,\bk)$ as defined in \cite{Babiarz:2020jkh} and $\tilde \psi^*_{\sigma \bar \sigma} (z,k_\perp)$:
\begin{eqnarray}
    \tilde \psi^*_{++} (z,k_\perp) \equiv \frac{k_\perp}{\sqrt{z(1-z)}} \psi(z, k_\perp)\,, \quad \tilde \psi^*_{+-} (z,k_\perp) = \tilde \psi^*_{-+} (z,k_\perp) \equiv \frac{m_f(1-2z)}{\sqrt{z(1-z)}} \psi(z, k_\perp)\, .
\end{eqnarray}
%%%%
In particular, radiative decay width can be found from the relation:
\begin{equation}
 \Gamma_{\gamma \gamma}(\chi_{c0})  = \frac{\pi \alpha_{em}^2}{M_{\chi_{c0}}}\, |F_{\rm TT}(0)|^2\, ,
\end{equation}
where we take $M_{\chi_{c0}} = 3.41\, {\rm GeV}$ for the meson mass \cite{PDG_Workman:2022ynf}. We recall some well-known relations from the early years of quarkonium physics (see the review \cite{Novikov:1977dq} and references therein). Namely in the NR limit we obtain 
%%%\
\begin{eqnarray}
   \Gamma_{\gamma \gamma}(\chi_{c0})  = \alpha^2_{\rm em} e_f^4 \, \frac{2^4 \cdot 9 \cdot N_c}{M^4} \, |R'(0)|^2 \, , 
\end{eqnarray}
%%%
and therefore
%%%
\begin{eqnarray}
    {\cal R}  \equiv \frac{\Gamma_{\gamma \gamma}(^3 P_2)}{\Gamma_{\gamma \gamma}(^3 P_0)} = \frac{4}{15} \simeq 0.27 \, .
\end{eqnarray}
\begin{table}[h]
    \caption{Radiative decay widths obtained in the LFWF approach and the ratio ${\cal R} = \Gamma_{\gamma \gamma}(\chi_{c2}) / \Gamma_{\gamma \gamma}(\chi_{c0})$. }
    \centering
    \begin{tabular}{lccc}
    \hline
    \hline
      &  $\Gamma_{\gamma \gamma}(\chi_{c0})$ [keV] &  $\Gamma_{\gamma \gamma}(\chi_{c2})$ [keV]  & ${\cal R} = \frac{\Gamma_{\gamma \gamma}(\chi_{c2})}{\Gamma_{\gamma \gamma}(\chi_{c0})}$\\
     \hline
     Cornell                    & 0.44 & 0.15 & 0.34\\
     logarithmic                & 0.91 & 0.32 & 0.35\\
     Buchm{\"u}ller-Tye         & 0.96 & 0.33 & 0.34\\
     power-like                 & 1.32 & 0.46 & 0.35\\
     harmonic oscillator        & 0.98 & 0.33 &  0.34\\
     BLFQ  N8                   & 1.70 & 0.39 &  0.23 \\
     BLFQ  N16                  & 2.03 & 0.41 &  0.20 \\
     BLFQ  N24                  & 2.18 & 0.40 &  0.18   \\
     BLFQ  N32                  & 2.33 & 0.40 & 0.17\\
     \hline
     PDG \cite{PDG_Workman:2022ynf} & 2.20 $\pm$ 0.15 & 0.56 $\pm$ 0.03 & 0.25 $\pm$ 0.02\\
     BES III    \cite{BESIII:2017rpg} &$2.03 \pm 0.08 \pm 0.06 \pm 0.13$ &  $0.60 \pm 0.02 \pm 0.01 \pm 0.04$ &  $0.295 \pm 0.014 \pm 0.007 \pm 0.027$ \\
     Belle \cite{Belle:2022exn}&  &$0.653 \pm 0.013\pm 0.031 \pm 0.017$  & \\
     CLEO \cite{CLEO:PhysRevD.78.09150} & $2.36 \pm 0.35 \pm 0.22$ & $0.66 \pm 0.07 \pm 0.06$ & $0.278 \pm 0.050 \pm 0.036$ \\
     \hline
     \hline
    \end{tabular}
    \label{tab:Ratio_Gamma}
\end{table}

In Tab.~\ref{tab:Ratio_Gamma}, we present results for $\Gamma_{\gamma \gamma}(\chi_{c0})$, $\Gamma_{\gamma \gamma}(\chi_{c2})$ and for their ratio (last column) for different $c \bar c$ potentials. In contrast to individual widths we get rather stable ratio $\Gamma_{\gamma \gamma}(\chi_{c2})   / \Gamma_{\gamma \gamma}(\chi_{c0} )\sim 0.34-0.35$. For comparison using BLFQ wave functions with different numbers of expansion terms (N8, N16, N24, N32) from the database \cite{Li:PRD}. In this case, the ratio is significantly smaller. For completeness, we also collected experimental results from the BESIII, Belle, and CLEO collaborations.

%%%%----------------------
\section{$\gamma^* \gamma$ cross-section and off-shell width}
\label{app:cross_sec}
%%%%-------------------------

Now, we wish to compare the $Q^2$-dependence of our form factors to the sparse data available from single-tag experiments.

The definition of off-shell widths that we were using comes from writing the $\gamma^* \gamma$ cross-section for photons as ($i,j \in T, L$) \cite{Olsson:1987jk}
%%%%
\begin{eqnarray}
    \sigma_{ij} &=& \frac{32 \pi}{N_i N_j} (2 J+1) \frac{W^2}{2 \sqrt{X}} \frac{\Gamma \Gamma^*_{ij}(Q^2)}{(W^2 - M^2)^2 + M^2 \Gamma^2} \nonumber \\
    &=& \frac{32 \pi}{N_i N_j} (2 J+1) \frac{W^2}{2 M\sqrt{X}} \, {\rm BW}(W^2,M^2) \, \Gamma^*_{ij}(Q^2)\, .    \label{eq:def_width_off_shell}
\end{eqnarray}
%%%%
For the case of one off-shell photon, we have that the kinematical factor $\sqrt{X} = \half (M^2 + Q^2)$.
Further, $N_T=2, N_L = 1$, and $J$ is the spin of the resonance of mass $M$ and total decay width $\Gamma$. By $\rm BW(W^2, M^2)$ we denote the Breit-Wigner distribution, which in the narrow width limit becomes
%%%%
\begin{eqnarray}
    {\rm BW}(W^2,M^2) \to \frac{\pi}{2M} \delta(W-M) \, . 
\end{eqnarray}

%%%%
Now, the $\rm TT$ and $\rm LT$ cross sections are obtained from the c.m.-frame helicity amplitudes as \cite{Budnev:1975poe}
%%%
\begin{eqnarray}
    \sigma_{\rm TT} &=& \frac{1}{4 \sqrt{X}} \Big( {\cal M}^*(++) {\cal M}(++) + {\cal M}^*(+-) {\cal M}(+-) \Big) \, {\rm BW}(W^2, M^2)\,, \nonumber \\
     \sigma_{\rm LT} &=& \frac{1}{2 \sqrt{X}} \, {\cal M}^*(0+) {\cal M}(0+) \, {\rm BW}(W^2, M^2) \, .
\end{eqnarray}
%%%%%%
%%%%
\begin{figure}[h!]
    \centering
    \includegraphics[width = 0.49 \textwidth]{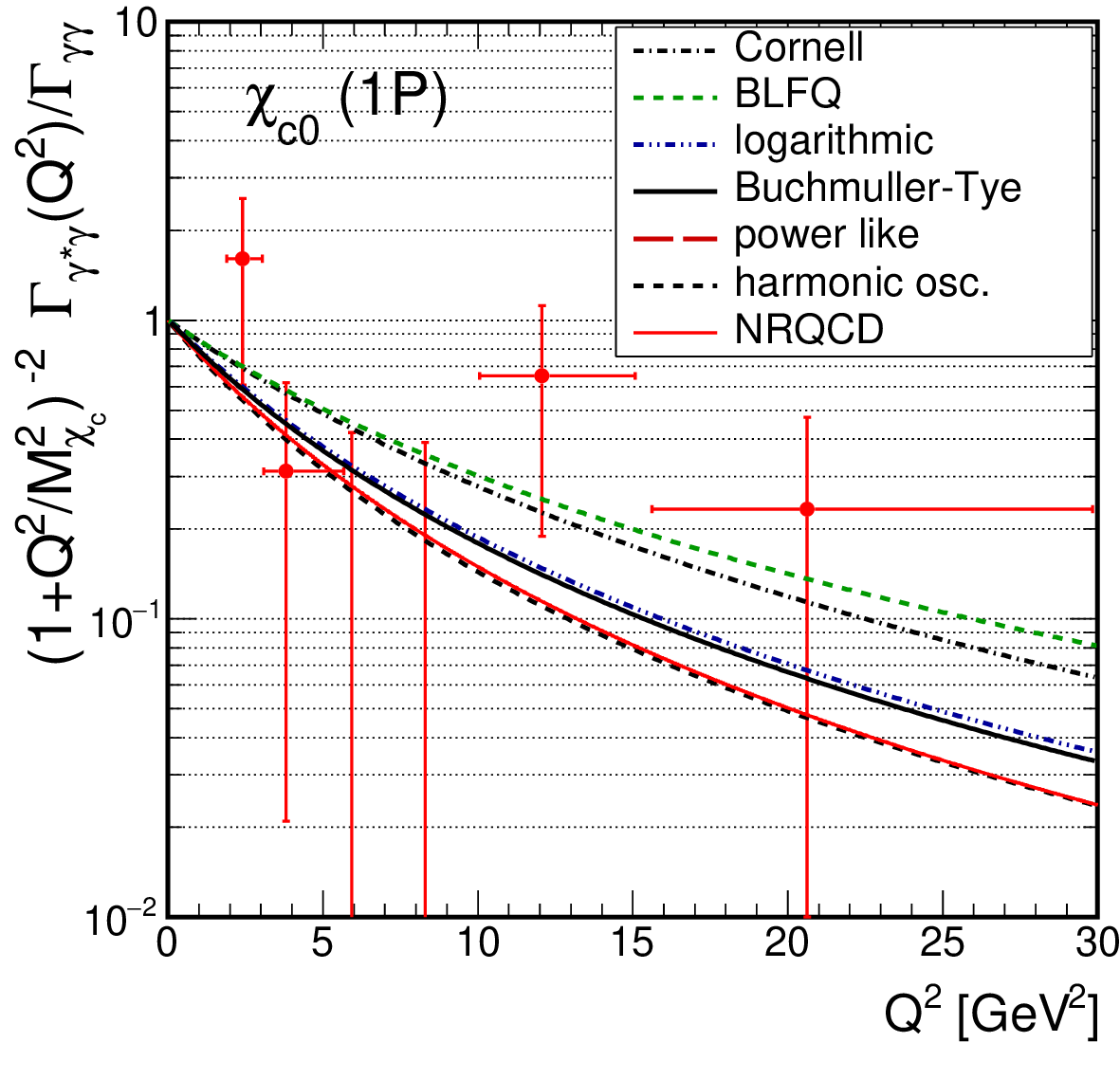}
    \includegraphics[width = 0.49\textwidth]{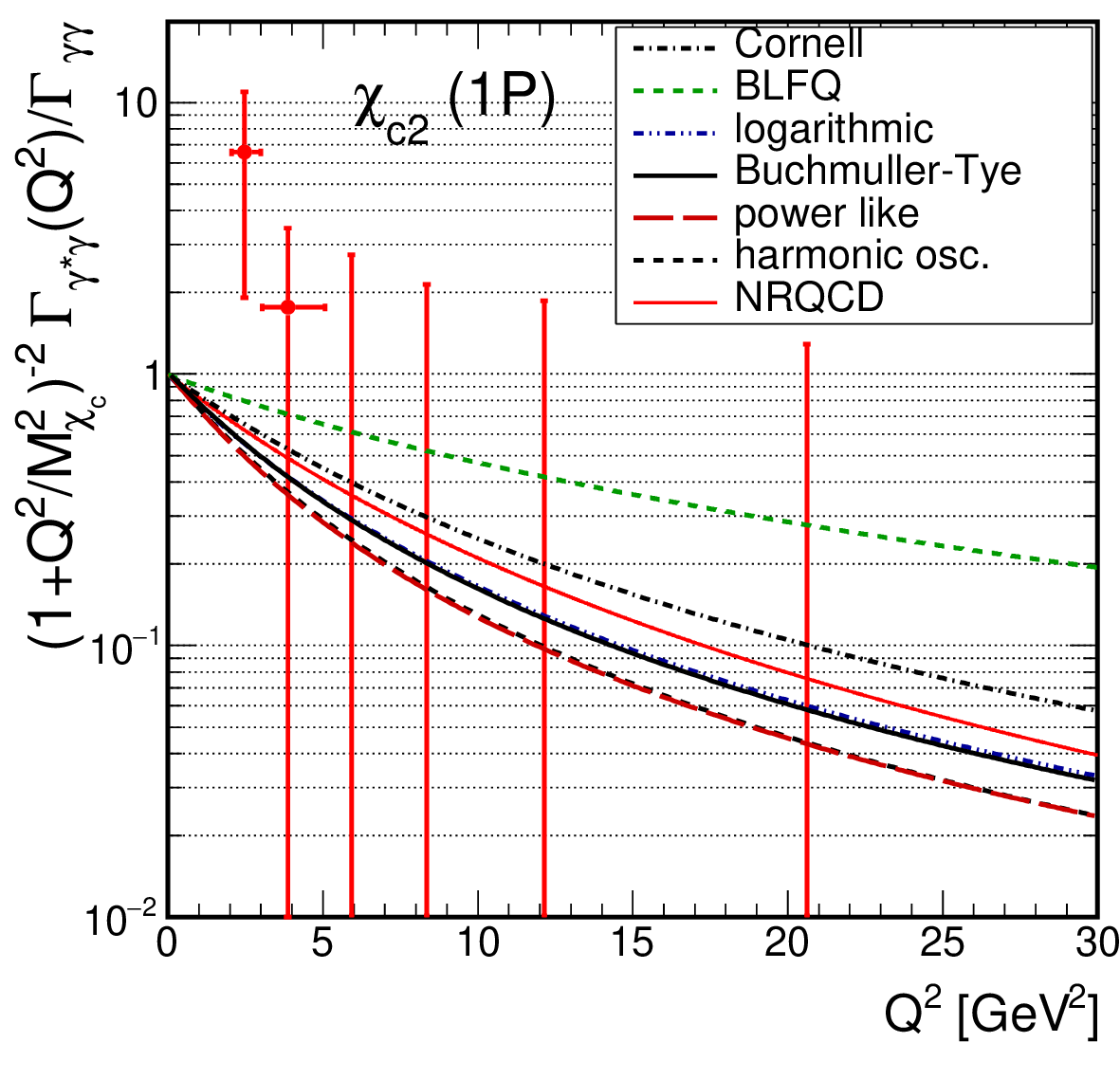}
    \caption{Off-shell decay width $\Gamma^*(Q^2)$ for $\chi_{c0}$ (on the l.h.s.) and $\chi_{c2}$ (on the r.h.s.) compared to the Belle data \cite{Belle:2017xsz}. In the NRQCD approach, we took $ M = M_{\chi_c}.$ }
    \label{fig:Gamma_Q2}
\end{figure}
%%%%%
Using the formulas in Ref.~\cite{Poppe:1986dq}, we relate our FFs to the helicity amplitudes, and obtain for the $\rm TT$ case:
%%%%
\begin{eqnarray}
    \sigma_{\rm TT}
    &=&\frac{(4 \pi \alpha_{\rm em})^2}{4 \sqrt{X}} 
    \Big\{  F^2_{\rm TT,2}(Q^2) + \frac{2}{3} \Big( 1 + \frac{Q^2}{M^2} \Big)^4 \, M^4 \, F^2_{\rm TT,0}(Q^2) \Big\} {\rm BW}(W^2,M^2)  \, .
\end{eqnarray}
%%%%%
and, for ${\rm LT}$:
%%%%
\begin{eqnarray}
    \sigma_{\rm LT} = \frac{Q^2 \sqrt{X}}{W^2} \, (4 \pi \alpha_{\rm em})^2 \,  F^2_{\rm LT}(Q^2) \, BW(W^2,M^2) \, ,
\end{eqnarray}
%%%%%%%%%
Comparing to Eq.~(\ref{eq:def_width_off_shell}), with $N_T = 2, J=2, W=M$, we derive the off-shell widths
%%%%%
\begin{eqnarray}
    \Gamma^*_{\rm TT}(Q^2) 
    &=& (4 \pi \alpha_{\rm em})^2 \Big\{ \frac{F^2_{\rm TT2}(Q^2)}{80 \pi M} + \frac{M^3 F^2_{\rm TT 0}(Q^2)}{120 \pi} \Big( 1 + \frac{Q^2}{M^2} \Big)^4  \, \Big\} \, .
\end{eqnarray}
%%%%%
For $Q^2=0$ this agrees with the formula for the two-photon decay width; see Eq.~(\ref{eq:Gamma_FF}).
%%%%
For the $\rm LT$ case, we obtain
%%%%
\begin{eqnarray}
    \Gamma^*_{\rm LT}(Q^2) =(4 \pi \alpha_{\rm em})^2 \frac{1}{160 \pi} \Big( 1 + \frac{Q^2}{M^2} \Big)^2 M Q^2 F^2_{\rm LT}(Q^2) \, .  \end{eqnarray}
%%%%
Let us now turn to the $Q^2$-dependence of the single-tag cross-section, which we write as:
%%%%
\begin{eqnarray}
    \frac{d \sigma}{dQ^2} = 2 \, \int dW \frac{d L}{dW dQ^2} \, \Big( \sigma_{\rm TT}(W^2,Q^2) + \epsilon_0 \sigma_{\rm LT}(W^2,Q^2) \Big) \, . 
\end{eqnarray}
%%%
The factor two appears because each of the lepton can emit the off-shell photon.
%%%%
In the narrow-width approximation, we therefore have
%%%%
\begin{eqnarray}
  \frac{d \sigma}{dQ^2} = 4 \pi^2 \frac{(2J +1)}{M^2} \Big(1 + \frac{Q^2}{M^2}\Big)^{-1} \frac{2 \, dL}{dWdQ^2}\Big|_{W=M} \, \Gamma_{\gamma^* \gamma}(Q^2) \, ,
\end{eqnarray}
%%%
with the effective off-shell width defined as
%%%%
\begin{eqnarray}
   \Gamma_{\gamma^* \gamma}(Q^2) =  \Gamma^*_{\rm TT}(Q^2) + \epsilon_0 2 \Gamma^*_{\rm LT}(Q^2) \, .
\end{eqnarray}

%%%%
Off-shell widths are convention-dependent, and to compare to the experimental data from Ref.~\cite{Belle:2017xsz}, we note that the Belle collaboration writes
%%%%
\begin{eqnarray}
  \frac{d \sigma}{dQ^2} = 4 \pi^2 \frac{(2J +1)}{M^2} \Big(1 + \frac{Q^2}{M^2}\Big) \frac{2 \, dL}{dWdQ^2}\Big|_{W=M} \, \Gamma^{\rm Belle}_{\rm \gamma^* \gamma}(Q^2) \, ,
\end{eqnarray}
%%%%%
which means, that
%%%%
\begin{eqnarray}
 \Gamma^{\rm Belle}_{\gamma^* \gamma}(Q^2) = \Big(1 + \frac{Q^2}{M^2}\Big)^{-2} \, \, \Gamma_{\gamma^* \gamma}(Q^2) \, .  
\end{eqnarray}
Then the cross-section for $\chi_{c2}$ can be written as:
\begin{eqnarray}
    \frac{d\sigma}{dQ^2} = 4 \pi^2 \frac{ (2J +1)}{M^2} \Big( 1 +\frac{Q^2}{M^2}\Big)^{-1} \frac{2dL}{dWdQ^2}\Big|_{W=M} \Big(\Gamma^*_{\rm TT}(Q^2) + \epsilon_0 2 \Gamma^*_{\rm LT}(Q^2) \Big)\, .
\end{eqnarray}
In the case of $\chi_{c0}$ we have only one form factor, which has transverse contribution $F_{\rm TT}$. According to Ref.~\cite{Poppe:1986dq} the cross-section for scalar meson has the form:
\begin{eqnarray}
    \sigma_{\rm TT} = \frac{(4 \pi \alpha_{\rm em})^2}{4\sqrt{X}} F^2_{\rm TT}(Q^2) \, .
\end{eqnarray}
Therefore, the off-shell width for $\chi_{c0}$ is:
\begin{eqnarray}
    \Gamma^*(Q^2) = \frac{(4 \pi \alpha_{\rm em})^2}{16 \pi M} F^2_{\rm TT}(Q^2)\,.
\end{eqnarray}

In Fig.~\ref{fig:Gamma_Q2} we present the off-shell decay width for $\chi_{c0}$ (l.h.s.) and $\chi_{c2}$ (r.h.s.). We show explicit factor $\Big( 1+\frac{Q^2}{M^2}\Big)^{-2}$ on the y-axis caption due to the difference between our definition and the one used by the Belle collaboration. The existing data are not sufficient to judge which potential model works better. Future Belle data could provide valuable information on this issue.

%--------------------------
\section{Conclusions}
%--------------------------

In the present paper, we have extended our light-front formulation to a formalism of 
photon transition form factors to the case of  $\gamma \gamma^* \to 2^{++}$ couplings (helicity
form factors) in terms of the light-front quark-antiquark wave functions of the meson.
We have presented detailed formulae for $F_{\rm TT,0}$, $F_{\rm TT,2}$
as well as $F_{\rm LT}$ form factors expressed in terms of the
light-front wave functions.
To obtain light-front wave functions, we use methods discussed previously in Ref.~\cite{Babiarz:2020jkh} for five different $c\bar c$ potential models, see also Appendix \ref{app:LFWF_melosh}. In addition, we have used the light-front wave functions from the Basis Light Front Quantization approach of \cite{Li:PRD, Li:2021ejv}.

The two-photon decay width is smaller than the value measured
by the Belle collaboration. This can be caused by too approximate
$c \bar c$ wave functions and/or higher Fock components in the 
$\chi_{c2}$ wave function and requires further studies which go beyond the scope of
the present letter.
We find the $\Gamma(\lambda = 0)/\Gamma(\lambda = \pm 2)$ ratio of the order of $10^{-3}$, which 
is in agreement with the current experimental precision.

We also have shown helicity form factor results 
for one real and one virtual photon as a function of the photon
virtuality. We have obtained a large spread of the results
for different potentials. The form factor results are ready to 
be verified e.g. by the Belle collaboration in single-tag $e^+ e^-$ collisions.
Furthermore, we have defined and calculated the so-called $Q^2$-dependent off-shell diphoton width and compared it to the Belle data. It is rather difficult to conclude on the consistency of the model with rather low statistics of the available Belle data.

\section*{Acknowledgements}
This work was partially supported by the Polish National Science Center grant UMO-2018/31/B/ST2/03537 and by the Center for Innovation and Transfer of Natural Sciences and Engineering Knowledge in Rzesz{\'o}w. R.P.~is supported in part by the Swedish Research Council grant, contract number 2016-05996, as well as by the European Research Council (ERC) under the European Union's Horizon 2020 research and innovation program (grant agreement No 668679).

%%%%%%%%%
\appendix
%%%%%%%%%
\section{LFWFs in Melosh transform approach}
\label{app:LFWF_melosh}
%%%%%%%%%
%%%%%%%%
We first need the Melosh transform of the operator ${\cal O} = \vec \sigma \cdot \vec e$, which is defined as:
%%%
\begin{eqnarray}
    {\cal O}' = R^\dagger(z,\vec k_\perp) {\cal O} R(1-z,\vec k_\perp) \, ,
\end{eqnarray}
%%%
see e.g. \cite{Babiarz:2020jkh} for an explicit definition of $R$.
%%%
Using the identity 
%%%%
\begin{eqnarray}
(\vec \sigma \cdot \vec a)(\vec \sigma \cdot \vec b)(\vec \sigma \cdot \vec a) = 2 (\vec a \cdot \vec b) (\vec \sigma \cdot \vec a) - \vec a^2 (\vec \sigma \cdot \vec b) \, ,
\end{eqnarray}
%%%%
we obtain, using our master formula \cite{Babiarz:2020jkh} that
%%%%
\begin{eqnarray}
{\cal O}' &=& \frac{1}{\sqrt{z(1-z)}} \frac{1}{M_0(M_0+2 m_f)} \Big\{ \vec \sigma \cdot \vec e \, 
\Big( 2 z(1-z) M_0^2 + M_0 m_f \Big)  \nonumber \\
&& - 2 \vec e \cdot (\vec n \times \vec k) \, \vec \sigma \cdot (\vec n \times \vec k) 
 + i (M_0 + 2m_f) \vec e\cdot (\vec n \times \vec k) \, \openone \nonumber \\
&& -  (2z-1) M_0 \Big( \vec e \cdot \vec k \, \vec \sigma \cdot \vec n - \vec e \cdot \vec n \, \vec \sigma \cdot \vec k \Big) \Big\} \, . 
\end{eqnarray}
%%%%
Notice, that in this section, $M_0$ denotes the invariant mass of the $Q \bar Q$ pair, i.e.
%%%%
\begin{eqnarray}
    M^2_{0} = \frac{\vec k_\perp^2 + m_f^2}{z(1-z)} \, .
\end{eqnarray}

Here, it is be useful to simplify
%%%
\begin{eqnarray}
    2 z(1-z) M_0^2 + M_0 m_f 
    %&= &\half \Big(4 z(1-z)M_0^2 + 2 M_0 m_f \Big) % + 2 M_0 m_f - (1-2z)^2 M_0^2 \Big) 
    %\nonumber \\
    &=& \half \Big( M_0 (M_0 +2m_f) - (1-2z)^2 M_0^2 \Big)\, .
\end{eqnarray}

The polarization vector can now be either longitudinal $\vec e = \vec n$, or transverse, $\vec e = \vec e_\perp$. Some simplifications occur in either case.
Let us start with the longitudinal case:
%%%
\begin{eqnarray}
    {\cal O}'_0 = \frac{1}{\sqrt{z(1-z)}} \, \Big\{ \vec \sigma \cdot \vec n \, \half \Big(
    1 - \frac{(2z-1)^2 M_0}{M_0+2m_f} \Big) + (2z-1) \frac{\vec \sigma \cdot \vec k_\perp}{M_0+2m_f} \Big\} \, .
\end{eqnarray}
%%%
Now, we need the vertex
%%%
\begin{eqnarray}
    \Gamma^{(0)}_{\sigma \bar \sigma} = {\cal O}'_0 \, i\sigma_2 \, , 
\end{eqnarray}
%%%
\begin{eqnarray}
\Gamma^0_{\sigma \bar \sigma} = \frac{1}{\sqrt{z(1-z)}} \Big\{
\begin{pmatrix}
       0 & 1 \\
       1  & 0 \\
\end{pmatrix}
\, \half \Big(
    1 - \frac{(2z-1)^2 M_0}{M_0+2m_f} \Big)
    + \frac{(2z-1) k_\perp}{M_0+2m_f} 
    \begin{pmatrix}
       - e^{-i \phi}& 0 \\
       0  & e^{i \phi} 
\end{pmatrix}
\Big\}\, .
\end{eqnarray}
%%%%
For the transverse polarization, we obtain
%%%%
\begin{multline}
{\cal O}'_\perp = \frac{1}{\sqrt{z(1-z)}} \frac{1}{M_0(M_0+2m_f)}\, \Big\{ \vec \sigma \cdot \vec e_\perp \Big(2z(1-z) M_0^2 + M_0 m_f \Big)\\
- 2 [\vec k_\perp, \vec e _\perp][\vec k_\perp, \vec \sigma_\perp] + i (M_0+2 m_f) [\vec k_\perp, \vec e_\perp] \openone  \\
- (2z-1) M_0 \vec e_\perp \cdot \vec k_\perp \, \vec \sigma \cdot \vec n \Big\}\, .
\end{multline}
%%%
Here, we have used that
\begin{eqnarray}
\vec a \cdot (\vec n \times \vec b) = \vec n \cdot (\vec b \times \vec a) = [\vec b _\perp, \vec a_\perp] = b_x a_y - b_y a_x    \, .
\end{eqnarray}
%%%
Furthermore, for
%%%
\begin{eqnarray}
    \vec e_\perp(\lambda) = - \frac{1}{\sqrt{2}} (\lambda \vec e_x + i \vec e_y )\, , 
    \label{eq:polar_vecT}
\end{eqnarray}
%%%%
we can write
%%%%
\begin{eqnarray}
 {\cal O}'_\perp &=&   \frac{1}{\sqrt{z(1-z)}} \frac{1}{M_0(M_0+2 m_f)}\Big\{ \vec \sigma \cdot \vec e_\perp \, m_f(M_0 + 2 m_f) - \sqrt{2} \lambda k_\perp e^{i \lambda \phi} \vec \sigma \cdot \vec k_\perp \nonumber \\
 &&+  (M_0 + 2 m_f) \frac{1}{\sqrt{2}} k_\perp e^{i \lambda \phi} \, \openone %\nonumber \\
  +(2z-1) M_0 \lambda \frac{1}{\sqrt{2}} k_\perp e^{i \lambda \phi} \vec \sigma \cdot \vec n \Big\} \nonumber \\
 &=& \frac{1}{\sqrt{z(1-z)}} \frac{1}{M_0} \Big\{ m_f \, \vec \sigma \cdot \vec e_\perp 
 + \frac{1}{\sqrt{2}} k_\perp e^{i \lambda \phi} \openone 
 - \frac {\sqrt{2} k_\perp\lambda}{M_0+2m_f} e^{i \lambda \phi}
 \vec \sigma \cdot \vec k_\perp 
 \nonumber \\
 &&+ \frac{(2z-1)M_0 k_\perp}{M_0 + 2m_f} \frac{\lambda}{\sqrt{2}} e^{i \lambda \phi} \, \vec \sigma \cdot \vec n \Big\} \, .\end{eqnarray}
%%%
Our vertex
%%%
\begin{eqnarray}
\Gamma^{(\lambda)}_{\sigma \bar \sigma} =  {\cal O}'_\perp \, i \sigma_2 \, ,
\end{eqnarray}
%%%
then becomes
%%%
\begin{eqnarray}
\Gamma^{(\lambda)}_{\sigma \bar \sigma} &=&  \frac{1}{\sqrt{z(1-z)}} \frac{1}{\sqrt{2}}\frac{1}{M}\Big\{
m_f
\begin{pmatrix}
       1+ \lambda & 0 \\
       0  & 1 - \lambda \\
\end{pmatrix} 
+ k_\perp
\begin{pmatrix}
       0  &  e^{i \lambda \phi} \\
       - e^{i \lambda \phi}  & 0 \\
       \end{pmatrix}
\nonumber \\
&-& \frac{2 k_\perp^2 \lambda}{M_0+ 2m_f}
\begin{pmatrix}
       - e^{i (\lambda -1) \phi}  &  0 \\
       0  & e^{i (\lambda +1) \phi} \\
\end{pmatrix} 
+ \frac{(2z-1) M_0 k_\perp}{M_0 + 2m_f} 
\begin{pmatrix}
       0 &  \lambda e^{i \lambda \phi} \\
       \lambda e^{i \lambda \phi}  & 0 \\
\end{pmatrix} \Big\} \, .
\end{eqnarray}
%%%%
Now we can construct the LFWF for the spin-2 state.
Namely, we start with the rest of the frame from the form:
%%%
\begin{eqnarray}
\hat \Psi^{\lambda}_{\tau \bar \tau}(\vec k) = \sqrt{\frac{3}{ 8\pi}} \,  \xi^\dagger_\tau  \, \sigma_i i\sigma_2 \xi_{\bar \tau} \, \frac{k_j}{k} \, E_{ij}(\lambda) \, \frac{u_1(k)}{k}\, ,
\end{eqnarray}
%%%
which satisfies the normalization condition
\begin{eqnarray}
  \sum_{\tau, \bar \tau} \int {\rm d}^3 \vec{k}\; \hat{\Psi}^{(\lambda)}_{\tau \bar \tau}(\vec k)\, \hat{\Psi}^{\dagger (\lambda')}_{\tau \bar \tau} (\vec k) = \delta_{\lambda \lambda'}\,  \qquad {\rm and} \qquad    \int u^2_1(k) {\rm d}k = 1 \, .
\end{eqnarray}
The polarization tensor is given by
%%%%
\begin{eqnarray}
    E_{ij}^{(\pm 2)} &=& e_i(\pm1) e_j (\pm 1)\, ,
    \nonumber \\
    E_{ij}^{(\pm1)} &=&  \frac{1}{\sqrt{2}} \Big( e_i(\pm1) n_j + n_i e_j(\pm 1) \Big)\, , \nonumber \\
    E^{(0)}_{ij} &=& \frac{1}{\sqrt{6}} \Big( e_i(+1) e_j(-1) + e_i(-1) e_j(+1) + 2 n_i n_j \Big) \, ,
    \label{eq:polar_tensor}
\end{eqnarray}
%%%%
where $\vec e (\lambda) = (\vec e_\perp(\lambda),0), \vec n = (0,0,1)$.
%%%%
Notice that the polarization tensor is symmetric and traceless,
\begin{eqnarray}
  E^{(\lambda)}_{ij} \delta_{ij} = 0 \, . 
\end{eqnarray}
%%%%%%
We have the operator ${\cal O}_{ij}$ for $^3P_2$ \cite{Gupta:1996ak}:
%%%%
\begin{eqnarray}
{\cal O}_{ij} = \sigma_i \, i \sigma_2 \frac{k_j}{k}\,,
\end{eqnarray}
%%%
where $k = \frac{1}{2}\sqrt{M^2_0 -4m_f^2}$, and
%%%
\begin{eqnarray}
   \hat{O}(\lambda) = \sqrt{\frac{3}{2}}\, {\cal O}_{ij} E^{(\lambda)}_{ij} \, , \quad {\rm Tr} [ \hat{O}(\lambda) \hat{O}^{\dagger}(\lambda)] = 1 \, ,
\end{eqnarray}
%%%
\begin{eqnarray}
\hat \Psi^{\lambda}_{\tau \bar \tau} =   \xi^\dagger_\tau  \hat{O}(\lambda) \xi_{\bar \tau }\frac{u_1(k)}{k} \sqrt{\frac{1}{ 4\pi}}\, .
\end{eqnarray}
%%%
Then, the vertex for the spin-2 meson is
%%%
\begin{eqnarray}
  \Phi^{(\pm 2)}_{\sigma \bar \sigma} &=& \mp \frac{1}{\sqrt{2}} \, \Gamma^{(\pm 1)}_{\sigma \bar \sigma} (k_x \pm i k_y)   = \frac{\mp1}{\sqrt{2}} \, \Gamma^{(\pm 1)}_{\sigma \bar \sigma} \, k_\perp e^{\pm i \phi}\, ,  \nonumber \\
\Phi^{(\pm 1)}_{\sigma \bar \sigma} &=&  {\frac{1}{ \sqrt{2}} } \Big( \Gamma^{(\pm 1)}_{\sigma \bar \sigma} (2z-1) {\frac{M}{2} } \mp \Gamma^{(0)}_{\sigma \bar \sigma} {\frac{k_\perp}{\sqrt{2}}} e^{\pm i \phi} \Big)\, , \nonumber \\
\Phi^{(0)}_{\sigma \bar \sigma} &=& \frac{1}{\sqrt{6}}
\Big(\Gamma^{(+ 1)}_{\sigma \bar \sigma} \frac{k_\perp}{\sqrt{2}} e^{-i \phi}
- \Gamma^{(- 1)}_{\sigma \bar \sigma} \frac{k_\perp}{\sqrt{2}} e^{i \phi} +  \Gamma^{(0)}_{\sigma \bar \sigma}  (2z-1) M \Big) \, .
\end{eqnarray}
%%%%
Then, the LFWF will have the form and normalization:
%%%%
\begin{eqnarray}
\Psi^{(\lambda)}_{\sigma \bar \sigma}(z,\vec k_\perp)  = \sqrt{\frac{3}{2}} \Phi^{(\lambda)}_{\sigma \bar \sigma} \, \phi(z,k_\perp)\, \frac{2}{\sqrt{M^2_0 -4m_f^2}} , \quad \int \frac{dz d^2\vec{k}_{\perp}}{z(1-z)16\pi^3} \sum_{\sigma, \bar \sigma} |\Psi^{(\lambda)}_{\sigma \bar \sigma}(z, k_{\perp})|^2 = 1 \, , \nonumber \\
\end{eqnarray}
\begin{eqnarray}
 \phi(z, k_{\perp}) = \sqrt{\frac{J}{4\pi}} \frac{u_1(k)}{k} = \pi \sqrt{M_0} \frac{u_1(k)}{k}\, , \quad \int \frac{dz d^2 \vec{k}_{\perp}}{ z(1-z) 16 \pi^3} |\phi(z, k_{\perp})|^2 = 1\,,  
\end{eqnarray}
%%%
\begin{multline}    
\Phi^{(0)}_{\sigma \bar \sigma} = \frac{1}{\sqrt{6 z(1-z)}}
%\times
\begin{pmatrix}
\frac{k_{\perp} e^{-i \varphi}}{M_0}\Big[m_f + \frac{2k^2_{\perp} -(2z-1)^2M_0^2}{M_0+2m_f})\Big] &  \quad (2z-1)\Big[\frac{k^2_{\perp}}{M_0+2m_f} + \frac{1}{2}M_0 - \frac{(2z-1)^2M_0^2}{2(M_0+2m_f)}\Big]\\
(2z-1)\Big[ \frac{k^2_{\perp}}{M_0 +2m_f}  +\frac{1}{2}M_0 -\frac{(2z-1)^2M_0^2}{2(M+2m_f)} \Big] & -\frac{k_{\perp}e^{i\varphi}}{M_0}\Big[ m_f + \frac{2k^2_{\perp} -(2z-1)^2M_0^2}{M_0 +2m_f}\Big] 
\end{pmatrix} \, ,
\end{multline}
%%%%
\begin{multline}
\Phi^{(+1)}_{\sigma \bar \sigma} = \frac{1}{2\sqrt{z(1-z)}}
\begin{pmatrix}
        m_f (2z-1) + 2k^2_{\perp}\frac{(2z-1)}{M_0+2m_f} & \quad k_{\perp}e^{i\varphi}\Big[ \frac{(2z-1)^2M_0}{M_0+2m_f} +(z-1)\Big]\\
         k_{\perp}e^{i\varphi}\Big[ \frac{(2z-1)^2 M_0}{M_0+2m_f} -z  \Big] & -2 k^2_{\perp}e^{i2\varphi}\frac{(2z-1)}{M_0+2m_f}
       \end{pmatrix}\, ,
\end{multline}
%%%
\begin{multline}
    \Phi^{(-1)}_{\sigma \bar \sigma} = \frac{1}{2\sqrt{z(1-z)}}
   \begin{pmatrix}
      -2k^2_{\perp} e^{-i2\varphi} \frac{(2z-1)}{M_0+2m_f} &  \quad -k_{\perp}e^{-i\varphi} \Big[ \frac{(2z-1)^2M_0}{M_0+2m_f}- z  \Big]\\
      -k_{\perp}e^{-i\varphi}\Big[ \frac{(2z-1)^2M_0}{M_0+2m_f} +(z-1) \Big] & m(2z-1)  + 2k^2_{\perp}\frac{(2z-1)}{M_0+2m_f}
    \end{pmatrix}\, ,
\end{multline}
%%%%
\begin{multline}    
\Phi^{(+2)}_{\sigma \bar \sigma} = \frac{-k_{\perp} e^{i\varphi }}{M_0\sqrt{ z(1-z)}}
\begin{pmatrix}
m_f+\frac{k^2_{\perp}}{M_0+2m_f} & \quad \frac{1}{2}k_{\perp} e^{i\varphi}(1+ \frac{(2z-1)M_0}{M_0+2m_f})  \\
-\frac{1}{2}k_{\perp} e^{i\varphi}(1- \frac{(2z-1)M_0}{M_0+2m_f})& -k^2_{\perp}e^{i2\varphi}\frac{1}{M_0+2m_f}
\end{pmatrix}\, ,
\end{multline}
%%%%
\begin{multline}    
\Phi^{(-2)}_{\sigma \bar \sigma} = \frac{k_{\perp} e^{-i\varphi }}{M_0\sqrt{ z(1-z)}}
\begin{pmatrix}
-k^2_{\perp} e^{-i2\varphi}\frac{1}{M_0+2m_f} & \quad \frac{1}{2}k_{\perp} e^{-i\varphi}(1- \frac{(2z-1)M_0}{M_0+2m_f})  \\
-\frac{1}{2}k_{\perp} e^{-i\varphi}(1+ \frac{(2z-1)M_0}{M_0+2m_f})& m_f+ \frac{k^2_{\perp}}{M_0+2m_f}
\end{pmatrix}\, .
\end{multline}
%%%

\bibliography{references.bib}

\end{document}